\documentclass[letterpaper,twocolumn,aps,prb,floatfix,amsmath,amssymb,nobibnotes,superscriptaddress,reprint]{revtex4-1}

\pdfoutput=1

\usepackage[latin9]{inputenc}
\usepackage{mathrsfs}
\usepackage{amsmath}
\usepackage{amssymb}
\usepackage{graphicx}
\usepackage{bm}
\usepackage{txfonts}
\usepackage{dcolumn}
\usepackage[caption=false,subrefformat=parens,labelformat=parens]{subfig}

\begin{document}

\title{A Green function method to study thin diffraction gratings }

\author{Daniel A. Travo}
\email{dtravo@physics.utoronto.ca}

\author{Rodrigo A. Muniz}

\affiliation{Department of Physics, University of Toronto, Toronto, Ontario M5S
1A7, Canada}

\author{M. Liscidini}

\affiliation{Department of Physics, University of Pavia, Pavia I-27100, Italy }

\author{J. E. Sipe}

\affiliation{Department of Physics, University of Toronto, Toronto, Ontario M5S
1A7, Canada}

\date{\today}
\begin{abstract}
The anomalous features in diffraction patterns first observed by Wood
over a century ago have been the subject of many investigations, both
experimental and theoretical. The sharp, narrow structures - and the
large resonances with which they are sometimes associated - arise
in numerous studies in optics and photonics. In this paper we present
an analytical method to study diffracted fields of optically thin
gratings that highlights the nonanalyticities associated with the
anomalies. Using this approach we can immediately derive diffracted
fields for any polarization in a compact notation. While our equations
are approximate, they fully respect energy conservation in the electromagnetic
field, and describe the large exchanges of energy between incident
and diffracted fields that can arise even for thin gratings. 
\end{abstract}
\maketitle

\section{Introduction }

Over a century ago, Wood observed anomalies in the angular dependence
of light reflected from a metal sheet \cite{Wood1902}, and since
then there have been many studies of these anomalies and their applications
in optics and photonics, particularly as they arise in the reflection
from surfaces on which a grating is intentionally deposited. Maystre
\cite{Maystre2012b} has recently presented a detailed review of the
history of research in this field. Generally, two mechanisms have
been identified as sources of the anomalies: The first is a transition
between propagation and evanescence in one of the diffracted orders
\cite{Rayleigh1907}, and the second is the excitation of a leaky
mode within the grating region \cite{Hessel1965}. Some ambiguity
exists in the literature concerning the distinction between these
mechanisms and the terms used to refer to them. In this paper, we
follow the convention that anomalies related to a transition in a
diffracted order are referred to as $Rayleigh$ $anomalies$, and
those associated with the resonant excitation of a leaky mode in the
grating region are referred to as $Wood$ $anomalies$. Maradudin
\cite{Maradudin2016} has recently shown that the resonant excitation
of any surface wave in a substrate below the grating, by scattering
from the grating, can lead to anomalies in the reflectance as well;
we also refer to these as Wood anomalies. 

Rayleigh anomalies are square root-like, sharp, narrow peaks that
arise in the irradiance of both the specularly reflected and diffracted
beams. Wood anomalies are associated with extraordinary increases
in the specular reflectance \cite{Falco2004,Zhao2010,Yi2016}, and
have seen wide use in applications where gratings serve as filters
\cite{Wang1995,Niederer2004,Brundrett1998}, modulators \cite{Grande2014,Grande2015},
and sensors \cite{Mukundan2009,Eitan2015}. Because even very thin
gratings can lead to very large effects on the reflectivity in the
region of these anomalies, the simplest perturbation theories are
not sufficient to describe them; it is essential to consider the full
interaction between diffracted and specularly reflected beams. Over
the years a wide range of approaches have arisen to treat such systems.
These include guided-mode techniques such as coupled mode theory \cite{Sipe2001,Yariv1973,Norton1997,Bykov2015},
transfer matrix approaches \cite{Wang1995}, and a variety of robust
numerical techniques based on finite element and RCWA (scattering
matrix) methods \cite{Li1996,Whittaker1999,Liscidini2008,Ahmed2010}. 

\begin{figure}
\centering

\subfloat[]{\includegraphics[width=0.75\columnwidth]{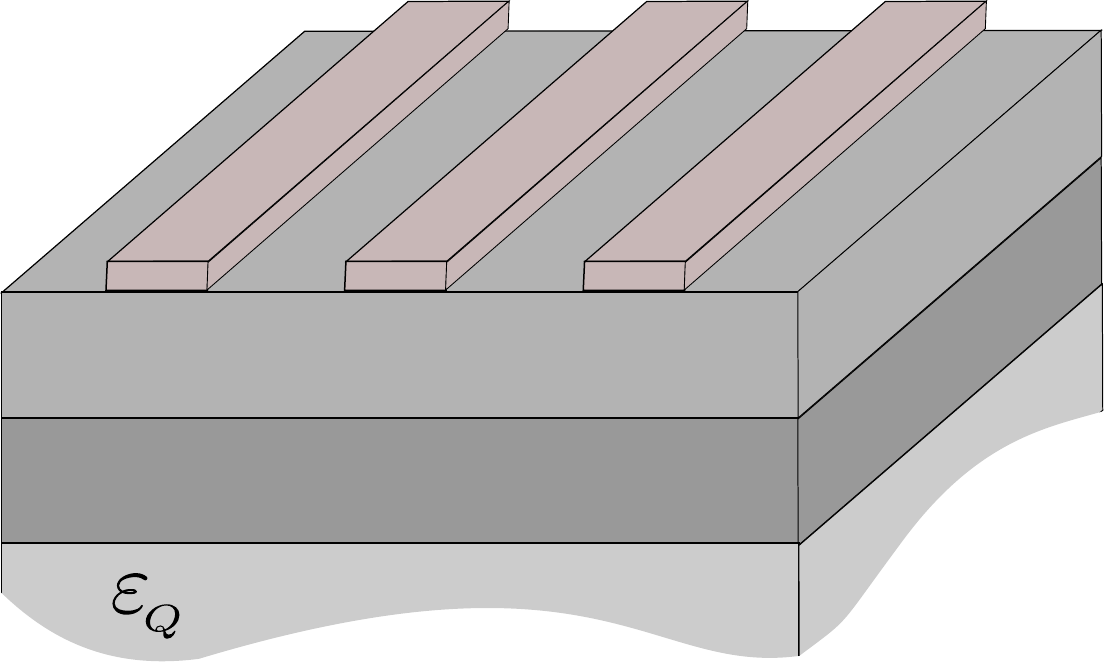}
\label{fig:grating_multi}}

\subfloat[]{\includegraphics[width=0.75\columnwidth]{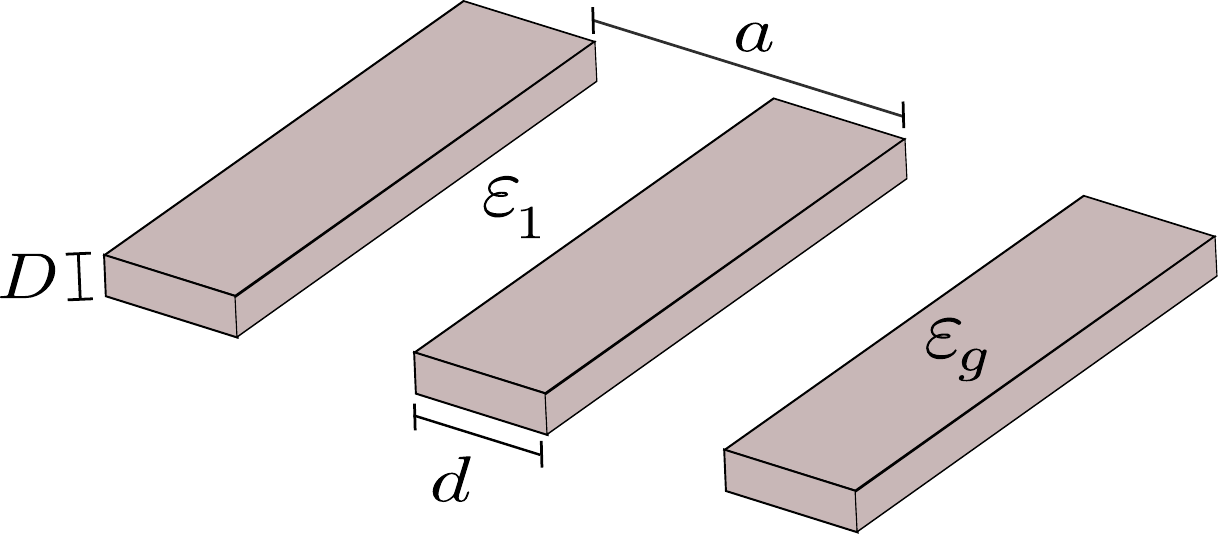}
\label{fig:grating_def}}

\subfloat[]{\includegraphics[width=0.75\columnwidth]{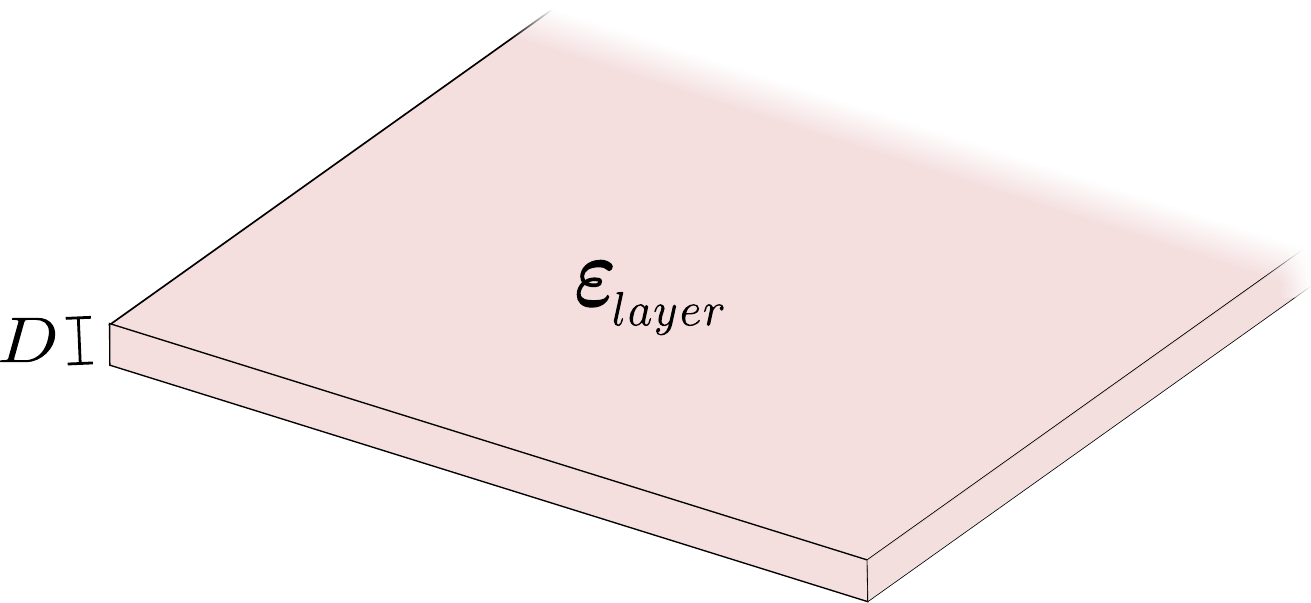}
\label{fig:eff_slab}}

\subfloat[]{\includegraphics[width=0.75\columnwidth]{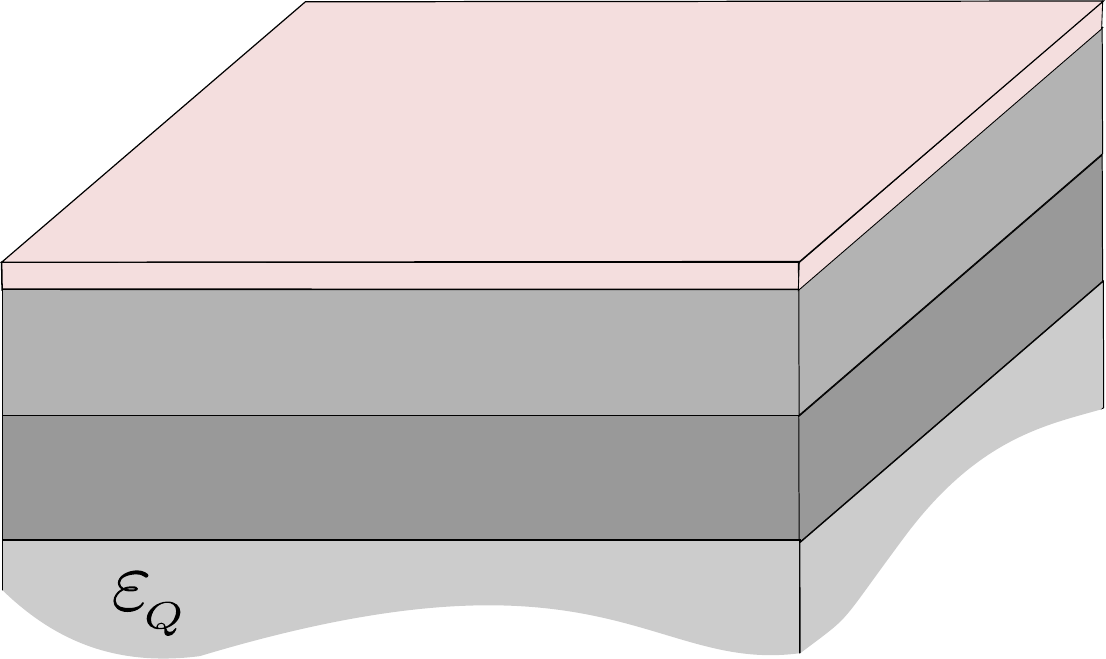}
\label{fig:eff_multi}}

\caption{(Color online) The general structure studied and discussed in this
paper (not to scale). (a) A thin grating placed on top of a multilayer
structure with a substrate with relative dielectric constant $\varepsilon_{Q}$.
The grating parameters are the same as in (b), and the relative dielectric
constant of the cladding is $\varepsilon_{1}$. (b) An isolated grating
with relative dielectric tensor $\varepsilon_{g}$ suspended in a
medium with relative dielectric constant $\varepsilon_{1}$. (c) An
effective dielectric slab with relative dielectric tensor $\boldsymbol{\varepsilon}_{layer}$
that characterizes the average properties of the grating relation.
(d) The corresponding effective planar structure consisting of the
effective dielectric slab and the multilayer structure.}
\end{figure}

In this paper we present a semi-analytic method for the treatment
of thin gratings, with advantages that are not all present in earlier
work. We consider the grating structure shown in Fig. \subref*{fig:grating_multi} in this
first communication. Based on a Green function formalism \cite{Sipe1987}
that treats the scattered light in terms of its $s-$ and $p-$polarized
components, the method leads to an immediate identification of the
features in the scattering equations that describe the anomalies,
and allows for the easy inclusion of effects of surface waves of the
substrate as well as leaky modes in the grating region. Light with
any plane of incidence, any polarization, and at any incident angle
is treated, and anisotropy in the response of the material in the
grating region is included; the substrate can consist of an arbitrary
set of layers with uniaxial optical properties. The description of
the reflected and diffracted light is necessarily approximate, since
we simplify our equations based on the grating being thin, but it
is nonetheless completely robust with respect to energy conservation:
In the absence of any absorption in the material media, at whatever
number the inclusion of diffracted and evanescent fields in the calculation
is truncated, the approximate equations respect conservation of energy
in the electromagnetic field, despite large exchanges of energy between
diffracted and reflected fields. For simple incidence configurations,
and if only a few diffracted orders are important, the set of equations
to be solved is small and the physics easily identified. This is an
important advantage, since gratings are now being used to access resonances
for enhanced sensing applications \cite{Liscidini2007,Liscidini2009}
and in novel 2D materials such as graphene \cite{Ju2011,Gao2012,Fu2017,Peng2017}.
Our simple but robust treatment of the optics of the grating should
allow for such work to focus on the physics of the medium being probed.

The outline of the paper is as follows. In section II we first treat
the simpler, symmetric grating structure shown in Fig.  \subref*{fig:grating_def}. In the
limit of a thin grating, we show how the scattering equations lead
naturally to the assignment of a uniform dielectric tensor for a layer
associated with the grating region; see Fig. \subref*{fig:eff_slab}. The scattering
by the grating can be best understood as occurring with this as part
of the background optical response, and it is the waveguide modes
of this nominal layer that become the resonances associated with the
Wood anomalies, discussed alongside Rayleigh anomalies in section
\ref{sec:Rayleigh-and-Wood} and identified in the resulting scattering
equations in section \ref{sec:Coupled-equations}. In section \ref{subsec:S-matrix-equations}
we build a scattering matrix for the problem. This can of course be
done in many ways, but we adopt an approach that leads to a proof
that the equations respect energy conservation, and allows for an
easy generalization to include an arbitrary layered substrate (Fig.
 \subref*{fig:grating_multi}). These equations are separated by polarization and simplified
in section \ref{subsec:simple-config} for a simple configuration
chosen as an example. In section \ref{subsec:An-example} a two wave-vector
model is used to derive analytic expressions for the scattered fields
alongside a discussion of their poles that signal the Wood anomalies.
We discuss how the Wood anomalies associated with the waveguide modes
of the grating region (Fig.  \subref*{fig:eff_slab}) are modified -- or disappear --
in the presence of the substrate in section \ref{subsec:substrate}
and present, as a sample calculation, results for a simple silicon
grating atop a glass substrate and confirm the validity of our approximate
treatment by comparison with convergent, numerically exact calculations.
Our conclusions are presented in section \ref{sec:Conclusions}. Some
of the details of the derivations, a discussion of waveguide dispersion,
and our proof of energy conservation are relegated to appendices. 

\section{Thin gratings \label{sec:Thin-gratings}}

We begin by considering a grating in the region $-D/2<z<D/2$, with
the rest of space taken to be filled with an isotropic dielectric;
see Fig.  \subref*{fig:grating_def}. In the presence of a field ${\bf E}_{in}\left({\bf r},t\right)$
incident on the grating region we write the total field as 
\begin{align}
\mathbf{E}\left(\mathbf{r},t\right) & =\mathbf{E}_{in}\left(\mathbf{r},t\right)+\mathbf{E}_{sc}\left(\mathbf{r},t\right),\label{eq:electric_field}
\end{align}
where $\mathbf{E}_{sc}\left(\mathbf{r},t\right)$ is the scattered
field. We take all such time dependent fields ${\bf F}\left({\bf r},t\right)$
to be stationary, 
\begin{align*}
\mathbf{F}\left({\bf r},t\right) & ={\bf F}\left({\bf r}\right)e^{-i\omega t}+c.c.,
\end{align*}
and we assume the refractive index of the surrounding dielectric,
$n_{1}=\sqrt{\varepsilon_{1}}$, is real at frequency $\omega$. Denoting
by $\mathbf{R}$ the projection in the $xy$ plane of a position vector
${\bf r}={\bf R}+z\hat{\boldsymbol{z}}$, we take $\hat{\boldsymbol{e}}$
to be the unit vector in the $xy$ plane that identifies the direction
in which the susceptibility varies, and write the (possibly complex)
spatially dependent (tensor) susceptibility in the grating region
as $\boldsymbol{\chi}\left(\zeta\right)$, where $\zeta=\hat{\boldsymbol{e}}\cdot\boldsymbol{\mathbf{R}}$. 

It will be convenient to Fourier transform our field amplitudes ${\bf F}\left({\bf r}\right)$
only in the $xy$ plane, 
\begin{align}
\mathbf{F}\left({\bf r}\right)=\mathbf{F}\left(\mathbf{R};z\right)= & \int\frac{d\boldsymbol{\kappa}}{\left(2\pi\right)^{2}}e^{i\boldsymbol{\kappa}\cdot{\bf R}}\mathbf{F}\left(\boldsymbol{\kappa};z\right),\label{eq:Fourier}
\end{align}
where $\mathbf{\boldsymbol{\kappa}}$ has $x$ and $y$ components,
so for example 
\begin{align}
\mathbf{E}(\boldsymbol{\mathbf{R}};z) & =\mathbf{E}_{in}(\mathbf{\boldsymbol{\mathbf{R}}};z)+\mathbf{E}_{sc}(\mathbf{\boldsymbol{\mathbf{R}}};z),\label{eq:Efull}\\
\mathbf{E}(\boldsymbol{\kappa};z) & =\mathbf{E}_{in}\left(\mathbf{\boldsymbol{\kappa}};z\right)+\mathbf{E}_{sc}\left(\mathbf{\boldsymbol{\kappa}};z\right).\nonumber 
\end{align}
In the usual example of an incident plane wave, the incident field
will be characterized by a single $\boldsymbol{\kappa}_{in}$, and
there will be scattered fields characterized by $\boldsymbol{\kappa}_{in}+m\mathbf{K}$,
where $m$ ranges over all positive and negative integers, and 
\begin{align}
\mathbf{K} & =\frac{2\pi}{a}\hat{\boldsymbol{e}},\label{eq:Kdef}
\end{align}
where $a$ is the fundamental period of the grating. For $\left|\boldsymbol{\kappa}_{in}+m\mathbf{K}\right|>\tilde{\omega}n_{1}$,
where $\tilde{\omega}\equiv\omega/c$, the scattered fields are evanescent,
confined to the neighborhood of the grating region; for $\left|\boldsymbol{\kappa}_{in}+m\mathbf{K}\right|<\tilde{\omega}n_{1}$
the scattering leads to diffracted fields that can carry energy away
from the grating region. 

We write 
\begin{align}
\boldsymbol{\chi}\left(\zeta\right) & =\boldsymbol{\chi}_{1}+\boldsymbol{\chi}_{add}\left(\zeta\right),\label{eq:Chidecomp}
\end{align}
where in dyadic form 
\begin{align*}
\boldsymbol{\chi}_{1} & =\left(\varepsilon_{1}-1\right)\left(\hat{\boldsymbol{x}}\hat{\boldsymbol{x}}+\hat{\boldsymbol{y}}\hat{\boldsymbol{y}}+\hat{\boldsymbol{z}}\hat{\boldsymbol{z}}\right),
\end{align*}
is just the susceptibility that would be present were the background
dielectric extended into the grating region, and $\boldsymbol{\chi}_{add}\left(\zeta\right)$
is an additional contribution that is responsible for the $\zeta$
dependence of the susceptibility in the grating region. We assume
that one of the principal axes of $\boldsymbol{\chi}_{add}\left(\zeta\right)$
is the $z$ axis, and we choose the $x$ and $y$ axes to coincide
with the other principal axes, 
\begin{equation}
\boldsymbol{\chi}_{add}\left(\zeta\right)=\hat{\boldsymbol{x}}\hat{\boldsymbol{x}}\chi_{add}^{xx}\left(\zeta\right)+\hat{\boldsymbol{y}}\hat{\boldsymbol{y}}\chi_{add}^{yy}\left(\zeta\right)+\hat{\boldsymbol{z}}\hat{\boldsymbol{z}}\chi_{add}^{zz}\left(\zeta\right).\label{eq:chi_add}
\end{equation}

We let $\mathbf{P}(\mathbf{r})$ denote the polarization in the grating
region, above and beyond that which would result if the grating region
consisted solely of the background isotropic dielectric medium. Then
\begin{align}
{\bf P}\left(\boldsymbol{\mathbf{R}};z\right) & =\epsilon_{0}\boldsymbol{\chi}_{add}\left(\zeta\right)\cdot{\bf E}\left(\boldsymbol{\mathbf{R}};z\right),\label{eq:polariz}\\
\text{for } & -D/2<z<D/2.\nonumber 
\end{align}
The scattered field contribution to (\ref{eq:Efull}) is determined
by 
\begin{equation}
{\bf E}_{sc}\left(\boldsymbol{\kappa};z\right)=\int_{-D/2}^{D/2}dz^{\prime}{\bf G}\left(\boldsymbol{\kappa};z-z^{\prime}\right)\cdot{\bf P}\left(\boldsymbol{\kappa};z^{\prime}\right),\label{eq:scat}
\end{equation}
where the Green function \cite{Sipe1987} is
\begin{align}
{\bf G}\left(\boldsymbol{\kappa};z-z^{\prime}\right) & ={\bf g}\left(\boldsymbol{\kappa};z-z^{\prime}\right)-\frac{1}{\epsilon_{0}\varepsilon_{1}}\delta\left(z-z^{\prime}\right)\hat{\boldsymbol{z}}\hat{\boldsymbol{z}},\label{eq:G_result}
\end{align}
with
\begin{align*}
{\bf g}\left(\boldsymbol{\kappa};z-z^{\prime}\right) & =\frac{i\tilde{\omega}^{2}}{2\epsilon_{0}w_{1}}\theta\left(z-z^{\prime}\right)e^{iw_{1}\left(z-z^{\prime}\right)}\left(\hat{\boldsymbol{s}}\hat{\boldsymbol{s}}+\hat{\boldsymbol{p}}_{1+}\hat{\boldsymbol{p}}_{1+}\right)\\
 & +\frac{i\tilde{\omega}^{2}}{2\epsilon_{0}w_{1}}\theta\left(z^{\prime}-z\right)e^{-iw_{1}\left(z-z^{\prime}\right)}\left(\hat{\boldsymbol{s}}\hat{\boldsymbol{s}}+\hat{\boldsymbol{p}}_{1-}\hat{\boldsymbol{p}}_{1-}\right),
\end{align*}
 and where
\begin{align}
\hat{\boldsymbol{s}} & =\hat{\boldsymbol{\kappa}}\times\hat{\boldsymbol{z}},\label{eq:unit_def}\\
\hat{\boldsymbol{p}}_{1\pm} & =\frac{\kappa\hat{\boldsymbol{z}}\mp w_{1}\hat{\boldsymbol{\kappa}}}{\tilde{\omega}n_{1}},\nonumber 
\end{align}
identify the $s-$ and $p-$polarized field components of the radiated
fields. Here $w_{1}\equiv\sqrt{\tilde{\omega}^{2}\varepsilon_{1}-\kappa^{2}}$,
where $\kappa=\left|\boldsymbol{\kappa}\right|$; to ensure proper
radiation conditions, the square root is made unique by taking ${\rm Im}\sqrt{Z}\geq0$
, and taking ${\rm Re}\sqrt{Z}\ge0$ if ${\rm Im}\sqrt{Z}=0$. 

Of the two terms on the right hand side of (\ref{eq:G_result}), the
second will typically lead to the larger contribution for our thin
gratings of interest, and it can be dealt with explicitly. If we define
a modified field, 
\begin{align}
\mathbf{E}_{mod}(\boldsymbol{\kappa};z)= & \mathbf{E}_{in}(\boldsymbol{\kappa};z)+\int_{-D/2}^{D/2}\mathbf{g}(\boldsymbol{\kappa};z-z')\cdot\mathbf{P}(\boldsymbol{\kappa};z')dz',\label{eq:Edr}
\end{align}
we have 
\begin{align*}
\mathbf{E}\left(\mathbf{R};z\right)= & \mathbf{E}_{mod}\left(\mathbf{R};z\right)-\frac{1}{\epsilon_{0}\varepsilon_{1}}\hat{\boldsymbol{z}}\hat{\boldsymbol{z}}\cdot\mathbf{P}(\boldsymbol{\mathbf{R}};z),
\end{align*}
and we can write the expression (\ref{eq:polariz}) for the polarization
as 
\begin{align}
{\bf P}\left(\boldsymbol{\mathbf{R}};z\right)= & \epsilon_{0}\boldsymbol{\chi}_{mod}\left(\zeta\right)\cdot{\bf E}_{mod}\left(\boldsymbol{\mathbf{R}};z\right),\label{eq:Pwork}
\end{align}
where 
\begin{align}
\boldsymbol{\chi}_{mod}\left(\zeta\right)= & \hat{\boldsymbol{x}}\hat{\boldsymbol{x}}\chi_{mod}^{xx}\left(\zeta\right)+\hat{\boldsymbol{y}}\hat{\boldsymbol{y}}\chi_{mod}^{yy}\left(\zeta\right)+\hat{\boldsymbol{z}}\hat{\boldsymbol{z}}\chi_{mod}^{zz}\left(\zeta\right),\label{eq:Chi_eff}
\end{align}
with 
\begin{align*}
\chi_{mod}^{xx}\left(\zeta\right) & \equiv\chi_{add}^{xx}\left(\zeta\right),\\
\chi_{mod}^{yy}\left(\zeta\right) & \equiv\chi_{add}^{yy}\left(\zeta\right),\\
\chi_{mod}^{zz}\left(\zeta\right) & \equiv\frac{\chi_{add}^{zz}\left(\zeta\right)}{1+\chi_{add}^{zz}\left(\zeta\right)/\varepsilon_{1}}.
\end{align*}
From (\ref{eq:Edr}) we see that as $D\rightarrow0$ we typically
have ${\bf E}_{mod}\left(\boldsymbol{\mathbf{R}};z\right)\rightarrow{\bf E}_{in}\left(\boldsymbol{\mathbf{R}};z\right)$,
and so in that limit $\boldsymbol{\chi}_{mod}\left(\zeta\right)$
can be understood as an effective local susceptibility relating the
(excess) polarization to the $incident$ $field$, rather than to
the field in the grating region itself.

So far we have made no approximations, and an exact description of
the scattering could proceed by numerically solving (\ref{eq:Edr},\ref{eq:Pwork})
for any specified $\boldsymbol{\mathbf{E}}_{in}(\boldsymbol{\kappa};z).$
Instead, we develop an approximate description of the scattering based
on the condition that the thickness $D$ of the grating region is
much less than the wavelength of light, $\tilde{\omega}n_{1}D\ll1$,
and as well that the variation in $z$ of the scattered fields over
the grating region is negligible for $\mathbf{\boldsymbol{\kappa}}$
of interest, $\left|w_{1}\right|D\ll1$. This leads to the $ansatz$
that a number of fields can be taken as independent of $z$ within
the grating region, 
\begin{align}
{\bf F}\left(\boldsymbol{\kappa};z\right) & \rightarrow{\bf F}\left(\boldsymbol{\kappa}\right)\label{eq:uniformity}\\
\text{for } & -D/2<z<D/2,\nonumber 
\end{align}
and for such fields we write 
\begin{align*}
\mathbf{F}(\mathbf{R})= & \int\frac{d\boldsymbol{\kappa}}{\left(2\pi\right)^{2}}e^{i\boldsymbol{\kappa}\cdot{\bf R}}\mathbf{F}\left(\boldsymbol{\kappa}\right).
\end{align*}
Naturally for the incident field we simply take $\mathbf{E}_{in}\left(\boldsymbol{\kappa}\right)=\mathbf{E}_{in}(\boldsymbol{\kappa};0)$,
while to determine $\mathbf{P}\left(\boldsymbol{\kappa}\right)$ self-consistently
we approximate the field $\mathbf{E}_{mod}(\boldsymbol{\kappa};z)$
as uniform over the grating region by taking $\mathbf{E}_{mod}(\boldsymbol{\kappa};z)\rightarrow\mathbf{E}_{mod}\left(\boldsymbol{\kappa}\right)$,
where 
\begin{align*}
\mathbf{E}_{mod}\left(\boldsymbol{\kappa}\right)= & \mathbf{E}_{in}\left(\boldsymbol{\kappa}\right)+\frac{1}{D}\int_{-D/2}^{D/2}dz\int_{-D/2}^{D/2}dz^{\prime}{\bf g}\left(\boldsymbol{\kappa};z-z^{\prime}\right)\cdot{\bf P}\left(\boldsymbol{\kappa}\right).
\end{align*}
In the limit $\left|w_{1}\right|D\ll1$ this leads to 
\begin{align}
\mathbf{E}_{mod}\left(\boldsymbol{\kappa}\right)= & \mathbf{E}_{in}\left(\boldsymbol{\kappa}\right)+{\bf g}\left(\boldsymbol{\kappa}\right)\cdot{\bf P}\left(\boldsymbol{\kappa}\right),\label{eq:scat0}
\end{align}
where 
\begin{align*}
{\bf g}\left(\boldsymbol{\kappa}\right) & =\frac{i\tilde{\omega}^{2}D}{2\epsilon_{0}w_{1}}\left[\hat{\boldsymbol{s}}\hat{\boldsymbol{s}}+\frac{1}{2}\left(\hat{\boldsymbol{p}}_{1+}\hat{\boldsymbol{p}}_{1+}+\hat{\boldsymbol{p}}_{1-}\hat{\boldsymbol{p}}_{1-}\right)\right],\\
 & =\frac{i\tilde{\omega}^{2}D}{2\epsilon_{0}w_{1}}\hat{\boldsymbol{s}}\hat{\boldsymbol{s}}+\frac{iw_{1}D}{2\epsilon_{0}\varepsilon_{1}}\hat{\boldsymbol{\kappa}}\hat{\boldsymbol{\kappa}}+\frac{i\kappa^{2}D}{2\epsilon_{0}\varepsilon_{1}w_{1}}\hat{\boldsymbol{z}}\hat{\boldsymbol{z}},
\end{align*}
and in this limit the equation (\ref{eq:Pwork}) reduces to 
\begin{align}
{\bf P}\left(\boldsymbol{\mathbf{R}}\right)= & \epsilon_{0}\boldsymbol{\chi}_{mod}\left(\zeta\right)\cdot{\bf E}_{mod}\left(\boldsymbol{\mathbf{R}}\right).\label{eq:Pwork2}
\end{align}
Within these approximations the fields in the grating region are determined
by the solution of (\ref{eq:scat0},\ref{eq:Pwork2}). 

At this point it is useful to separate out the spatial average of
our various quantities. In particular for $\boldsymbol{\chi}_{mod}\left(\zeta\right)$
we have 
\begin{align*}
\left\langle \boldsymbol{\chi}_{mod}\right\rangle = & \frac{1}{a}\int_{-a/2}^{a/2}\boldsymbol{\chi}_{mod}\left(\zeta\right)d\zeta,
\end{align*}
and we put 
\begin{align}
\boldsymbol{\chi}_{v}\left(\zeta\right)\equiv & \boldsymbol{\chi}_{mod}\left(\zeta\right)-\left\langle \boldsymbol{\chi}_{mod}\right\rangle .\label{eq:chiv_def}
\end{align}
Our equations (\ref{eq:scat0},\ref{eq:Pwork2}) can then be written
as 
\begin{align}
{\bf P}\left(\boldsymbol{\kappa}\right) & =\epsilon_{0}\left\langle \boldsymbol{\chi}_{mod}\right\rangle \cdot{\bf E}_{mod}\left(\boldsymbol{\kappa}\right)+{\bf P}_{v}\left(\boldsymbol{\kappa}\right),\label{eq:soe}\\
{\bf E}_{mod}\left(\boldsymbol{\kappa}\right) & ={\bf E}_{in}\left(\boldsymbol{\kappa}\right)+{\bf g}\left(\boldsymbol{\kappa}\right)\cdot{\bf P}\left(\boldsymbol{\kappa}\right),\nonumber 
\end{align}
where 
\begin{align}
{\bf P}_{v}\left(\boldsymbol{\mathbf{R}}\right)= & \epsilon_{0}\boldsymbol{\chi}_{v}\left(\zeta\right)\cdot{\bf E}_{mod}\left(\mathbf{R}\right)\label{eq:Pvdef}
\end{align}
is the only contribution from the variation of the effective susceptibility
with $\zeta$. If we define $\boldsymbol{\chi}_{layer}$ according
to 
\begin{align}
\chi_{layer}^{xx} & \equiv\left\langle \chi_{mod}^{xx}\right\rangle =\left\langle \chi_{add}^{xx}\right\rangle ,\nonumber \\
\chi_{layer}^{yy} & \equiv\left\langle \chi_{mod}^{yy}\right\rangle =\left\langle \chi_{add}^{yy}\right\rangle ,\label{eq:Chi_layer}\\
\frac{\chi_{layer}^{zz}}{1+\chi_{layer}^{zz}/\varepsilon_{1}} & \equiv\left\langle \chi_{mod}^{zz}\right\rangle =\left\langle \frac{\chi_{add}^{zz}}{1+\chi_{add}^{zz}/\varepsilon_{1}}\right\rangle ,\nonumber 
\end{align}
where the last equation is to be solved for $\chi_{layer}^{zz}$,
we can identify $\boldsymbol{\chi}_{layer}$ as the effective (excess)
susceptibility of the thin layer that would lead to the optical response
of the grating region were the variation $\boldsymbol{\chi}_{v}\left(\zeta\right)$
in the effective susceptibility ignored; this is the scenario sketched
in Fig.  \subref*{fig:eff_slab}. For if we would return to (\ref{eq:polariz},\ref{eq:scat}),
take $\boldsymbol{\chi}_{add}\left(\zeta\right)\rightarrow\boldsymbol{\chi}_{layer}$
and repeat the derivation and approximations leading to (\ref{eq:soe}),
we would recover precisely those equations with ${\bf P}_{v}\left(\boldsymbol{\kappa}\right)$
absent, and with the components of $\left\langle \boldsymbol{\chi}_{mod}\right\rangle $
replaced by the components of $\boldsymbol{\chi}_{layer}$ according
to (\ref{eq:Chi_layer}). Note that if we write the full relative
dielectric tensor in the grating region as $\boldsymbol{\varepsilon}\left(\zeta\right)\equiv\boldsymbol{\varepsilon}_{1}+\boldsymbol{\chi}_{add}\left(\zeta\right)$,
and the full relative dielectric tensor associated with $\boldsymbol{\chi}_{layer}$
as $\boldsymbol{\varepsilon}_{layer}\equiv\boldsymbol{\varepsilon}_{1}+\boldsymbol{\chi}_{layer}$,
where $\boldsymbol{\varepsilon}_{1}=\varepsilon_{1}(\hat{\boldsymbol{x}}\hat{\boldsymbol{x}}+\hat{\boldsymbol{y}}\hat{\boldsymbol{y}}+\hat{\boldsymbol{z}}\hat{\boldsymbol{z}})$,
we have 
\begin{align}
\varepsilon_{layer}^{xx} & =\left\langle \varepsilon^{xx}\right\rangle ,\nonumber \\
\varepsilon_{layer}^{yy} & =\left\langle \varepsilon^{yy}\right\rangle ,\label{eq:epsilon_layer}\\
\frac{1}{\varepsilon_{layer}^{zz}} & =\left\langle \frac{1}{\varepsilon^{zz}}\right\rangle .\nonumber 
\end{align}

We return to the equations (\ref{eq:soe}), and can now understand
them as describing the scattering due to a variation in the effective
excess susceptibility, $\boldsymbol{\chi}_{v}\left(\zeta\right)$,
in the presence of a uniform background dielectric tensor $\boldsymbol{\varepsilon}_{layer}$
in the grating region. Below we will construct an expression for ${\bf P}_{v}\left(\boldsymbol{\kappa}\right)$,
and then these equations can be solved consistently for ${\bf P}\left(\boldsymbol{\kappa}\right)$.
Once that is done we can construct the scattered fields above the
grating region ($z>D/2)$ and below the grating region ($z<-D/2)$.
We denote these by ${\bf E}_{sc}^{+}\left(\boldsymbol{\kappa};z\right)$
and ${\bf E}_{sc}^{-}\left(\boldsymbol{\kappa};z\right)$ respectively,
and they follow immediately from the general expression (\ref{eq:scat})
for the scattered field \cite{Sipe1987}; we have 
\begin{equation}
{\bf E}_{sc}^{\pm}\left(\boldsymbol{\kappa};z\right)=e^{\pm iw_{1}z}{\bf E}_{sc}^{\pm}\left(\boldsymbol{\kappa}\right),\label{eq:Escat}
\end{equation}
where
\begin{align}
{\bf E}_{sc}^{\pm}\left(\boldsymbol{\kappa}\right)= & {\bf G}^{\pm}\left(\boldsymbol{\kappa}\right)\cdot{\bf P}\left(\boldsymbol{\kappa}\right),\label{eq:Escat_def}
\end{align}
and 
\begin{equation}
{\bf G}^{\pm}\left(\boldsymbol{\kappa}\right)=\frac{i\tilde{\omega}^{2}D}{2\epsilon_{0}w_{1}}\left(\hat{\boldsymbol{s}}\hat{\boldsymbol{s}}+\hat{\boldsymbol{p}}_{1\pm}\hat{\boldsymbol{p}}_{1\pm}\right),\label{eq:Green}
\end{equation}
and we have again assumed $\left|w_{1}\right|D\ll1$. Now the incident
field satisfies the Maxwell equations with a uniform relative dielectric
constant $\varepsilon_{1}$, and so everywhere in space it is of the
form 
\begin{align}
\mathbf{E}_{in}(\boldsymbol{\kappa};z)= & \mathbf{E}_{in}^{+}(\boldsymbol{\kappa};z)+\mathbf{E}_{in}^{-}(\boldsymbol{\kappa};z),\label{eq:Ein}
\end{align}
where 
\begin{align}
{\bf E}_{in}^{\pm}\left(\boldsymbol{\kappa};z\right)= & e^{\pm iw_{1}z}{\bf E}_{in}^{\pm}\left(\boldsymbol{\kappa}\right),\label{eq:Ein_def}
\end{align}
($cf.$ (\ref{eq:Escat})). Then given any $\boldsymbol{\kappa}$
, for $z>D/2$ we label the full upward propagating (or evanescent)
fields as $\mathbf{E}_{out}^{+}\left(\boldsymbol{\kappa}\right)\exp(iw_{1}z)$,
while for $z<-D/2$ we label the full downward propagating (or evanescent)
fields as $\mathbf{E}_{out}^{-}\exp(-iw_{1}z$); we clearly have 
\begin{align}
\mathbf{E}_{out}^{\pm}\left(\boldsymbol{\kappa}\right)= & \mathbf{E}_{in}^{\pm}\left(\boldsymbol{\kappa}\right)+\mathbf{E}_{sc}^{\pm}\left(\boldsymbol{\kappa}\right).\label{eq:Eout_def}
\end{align}
Before solving for these fields, we identify how the Rayleigh and
Wood anomalies are captured in their calculation.

\section{Rayleigh and Wood Anomalies\label{sec:Rayleigh-and-Wood}}

Returning to the expression (\ref{eq:Green}) for ${\bf G}^{\pm}\left(\boldsymbol{\kappa}\right)$,
and writing $\hat{\boldsymbol{p}}_{1\pm}$ in terms of $\hat{\boldsymbol{\kappa}}$
and $\hat{\boldsymbol{z}}$, we see that in this basis of real unit
vectors there are terms in ${\bf G}^{\pm}\left(\boldsymbol{\kappa}\right)$
proportional to $w_{1}$, and terms proportional to $1/w_{1}$ These
are both non-analytic in $\kappa$, since $w_{1}$ is purely real
for $\kappa<\tilde{\omega}n_{1}$, purely imaginary for $\kappa>\tilde{\omega}n_{1}$,
and vanishes at $\kappa=\tilde{\omega}n_{1}$; $1/w_{1}$ thus diverges
at $\kappa=\tilde{\omega}n_{1}$. The transition from real to imaginary
$w_{1}$ can arise as the angle of incidence is varied, and $\boldsymbol{\kappa}$
is associated with a diffracted order that becomes evanescent in the
background dielectric as $\kappa$ first approaches, and then exceeds,
$\tilde{\omega}n_{1}$. Of course, although the ${\bf G}^{\pm}\left(\boldsymbol{\kappa}\right)$
diverge as $w_{1}\rightarrow0$, the ${\bf E}_{sc}^{\pm}\left(\boldsymbol{\kappa}\right)$
do not; the same non-analyticity as $w_{1}\rightarrow0$ appears in
${\bf g}\left(\boldsymbol{\kappa}\right)$, since
\begin{align*}
{\bf g}\left(\boldsymbol{\kappa}\right)= & \frac{1}{2}\left({\bf G}^{+}\left(\boldsymbol{\kappa}\right)+{\bf G}^{-}\left(\boldsymbol{\kappa}\right)\right),
\end{align*}
and once the expression for ${\bf P}_{v}\left(\boldsymbol{\kappa}\right)$
is included the self-consistent solution of the set of equations (\ref{eq:soe})
leads to finite fields everywhere at all $\kappa$, as we show in
detail below. This is enforced by the coupling among the different
diffracted and evanescent orders, and by the coupling between each
of them to the specularly reflected and transmitted fields; the source
of these couplings is of course the grating that is itself responsible
for the existence of the diffracted and evanescent orders themselves.
Another consequence of these couplings is that the non-analyticity
associated with the passing of a diffracted order into evanescence
appears as well in the expressions for the amplitudes of the other
diffracted orders, and in those of the specularly reflected and transmitted
fields. These are the Rayleigh anomalies. 

Another non-analyticity implicit in these equations can be revealed
by inserting the second of (\ref{eq:soe}) into the first and formally
solving for $\mathbf{P}\left(\boldsymbol{\kappa}\right)$,
\begin{align}
{\bf P}\left(\boldsymbol{\kappa}\right)= & \left(\mathbf{I}-\epsilon_{0}\left\langle \boldsymbol{\chi}_{mod}\right\rangle \cdot{\bf g}\left(\boldsymbol{\kappa}\right)\right)^{-1}\label{eq:EfromPg}\\
 & \cdot\left[\epsilon_{0}\left\langle \boldsymbol{\chi}_{mod}\right\rangle \cdot{\bf E}_{in}\left(\boldsymbol{\kappa}\right)+{\bf P}_{v}\left(\boldsymbol{\kappa}\right)\right],\nonumber 
\end{align}
where $\mathbf{I}=\hat{\boldsymbol{x}}\hat{\boldsymbol{x}}+\hat{\boldsymbol{y}}\hat{\boldsymbol{y}}+\hat{\boldsymbol{z}}\hat{\boldsymbol{z}}$
is the unit dyadic. The expression (\ref{eq:EfromPg}) is valid as
long as $\left(\mathbf{I}-\epsilon_{0}\left\langle \boldsymbol{\chi}_{mod}\right\rangle \cdot{\bf g}\left(\boldsymbol{\kappa}\right)\right)^{-1}$
has no divergent components, and this holds as long as the determinant
of a matrix representing $\left(\mathbf{I}-\epsilon_{0}\left\langle \boldsymbol{\chi}_{mod}\right\rangle \cdot{\bf g}\left(\boldsymbol{\kappa}\right)\right)$
does not vanish. In the special case where $\varepsilon_{layer}^{xx}=\varepsilon_{layer}^{yy}\equiv\varepsilon_{layer}^{\parallel}$
(recall (\ref{eq:Chi_layer},\ref{eq:epsilon_layer})), that matrix
can be easily written out in the $(\hat{\boldsymbol{s}},\hat{\boldsymbol{\kappa}},\hat{\boldsymbol{z}})$
basis, since $\hat{\boldsymbol{x}}\hat{\boldsymbol{x}}+\hat{\boldsymbol{y}}\hat{\boldsymbol{y}}=\hat{\boldsymbol{s}}\hat{\boldsymbol{s}}+\hat{\boldsymbol{\kappa}}\hat{\boldsymbol{\kappa}}$;
we find 
\begin{align}
 & \det\left(\mathbf{I}-\epsilon_{0}\left\langle \boldsymbol{\chi}_{mod}\right\rangle \cdot{\bf g}\left(\boldsymbol{\kappa}\right)\right)\label{eq:mode}\\
 & =\left[1-\frac{i\tilde{\omega}^{2}D}{2w_{1}}\left(\varepsilon_{layer}^{\parallel}-\varepsilon_{1}\right)\right]\left[1-\frac{iw_{1}D}{2\varepsilon_{1}}\left(\varepsilon_{layer}^{\parallel}-\varepsilon_{1}\right)\right]\nonumber \\
 & \times\left[1-\frac{i\kappa^{2}D}{2w_{1}}\frac{\varepsilon_{layer}^{\perp}-\varepsilon_{1}}{\varepsilon_{layer}^{\perp}}\right],\nonumber 
\end{align}
where we have put $\varepsilon_{layer}^{\perp}\equiv\varepsilon_{layer}^{zz}$.
For reasonable $\varepsilon_{layer}^{\parallel}$ the middle bracketed
term in (\ref{eq:mode}) cannot vanish, since by assumption $\left|w_{1}\right|D\ll1$;
thus the determinant in (\ref{eq:mode}) can vanish only if one of
the following conditions is met: 
\begin{align}
1-\frac{i\tilde{\omega}^{2}D}{2w_{1}}\left(\varepsilon_{layer}^{\parallel}-\varepsilon_{1}\right) & =0,\label{eq:mode_conditions}\\
1-\frac{i\kappa^{2}D}{2w_{1}}\frac{\varepsilon_{layer}^{\perp}-\varepsilon_{1}}{\varepsilon_{layer}^{\perp}} & =0.\nonumber 
\end{align}
In Appendix \ref{sec:WGmodes} we show that the first of (\ref{eq:mode_conditions})
is the dispersion relation for the fundamental $s$-polarized mode,
and the second for the fundamental $p$-polarized mode, of a thin
enough planar uniaxial waveguide with relative dielectric tensor susceptibility
$(\hat{\boldsymbol{x}}\hat{\boldsymbol{x}}+\hat{\boldsymbol{y}}\hat{\boldsymbol{y}})\varepsilon_{layer}^{\parallel}+\hat{\boldsymbol{z}}\hat{\boldsymbol{z}}\varepsilon_{layer}^{\perp}$,
bounded above and below by a uniform isotropic dielectric with dielectric
constant $\varepsilon_{1}$; recall that in the limit of a thin enough
planar waveguide at most one waveguide mode of each polarization exists.
Around the values of $\kappa$ where they vanish, the left-hand-sides
of (\ref{eq:mode_conditions}) can be written as proportional to $(\kappa-\kappa_{S})$
and $(\kappa-\kappa_{P})$ respectively, where at frequency $\omega$
the $s-$ and $p-$polarized waveguide modes have wave numbers $\kappa_{S}$
and $\kappa_{P}$ respectively; if there is no absorption, $\kappa_{S}$
and $\kappa_{P}$ are real. Thus the non-analyticities of $\left(\mathbf{I}-\epsilon_{0}\left\langle \boldsymbol{\chi}_{mod}\right\rangle \cdot{\bf g}\left(\boldsymbol{\kappa}\right)\right)^{-1}$are
poles, on the real $\kappa$ axis if there is no absorption, associated
with the waveguide modes of the ``effective waveguide'' established
by the average optical response in the grating region. 

Despite these divergences, the solution of (\ref{eq:EfromPg}) for
${\bf P}\left(\boldsymbol{\kappa}\right)$ is again always finite.
The waveguide modes exist for $\kappa>\tilde{\omega}n_{1},$ ``beyond
the light line,'' and no physical field incident from infinity can
be described by nonzero ${\bf E}_{in}\left(\boldsymbol{\kappa}\right)$
for $\kappa$ in the range of the divergences. Of course, by coupling
through the grating, ${\bf P}_{v}\left(\boldsymbol{\kappa}\right)$
can acquire $\boldsymbol{\kappa}$ components for $\kappa$ at wave
numbers near or at the waveguide modes if the angle of incidence of
the incident field is properly chosen, as we see in detail below.
However, a grating that allows ${\bf P}_{v}\left(\boldsymbol{\kappa}\right)$
to acquire those $\boldsymbol{\kappa}$ components from the incident
field will also couple part of any field that ${\bf P}_{v}\left(\boldsymbol{\kappa}\right)$
generates back to the wave vector of the incident field, thus modifying
the effective incident field driving ${\bf P}_{v}\left(\boldsymbol{\kappa}\right)$
and ameliorating the response; the effective waveguide pole is moved
off the real $\kappa$ axis, as we illustrate in an example later.
Another consequence is that the resonant structure associated with
one of the evanescent orders being close to an effective waveguide
mode will lead, through coupling by the grating, to resonant structures
in other diffracted and evanescent orders, and in the specularly reflected
and transmitted fields. These are the Wood anomalies.

Thus within the approximation of a thin grating region even a schematic
discussion as presented above can identify Rayleigh and Wood anomalies
with non-analyticities in the response of the grating structure to
an incident field: Rayleigh anomalies are associated with square root
divergences as a diffracted order becomes evanescent, and Wood anomalies
are associated with pole divergences as an evanescent order approaches
an effective waveguide mode of the grating region. Full calculations
within this approximation presented below will confirm this connection,
and show that our equations, while approximate, exhibit exact energy
conservation. As well, since for thin grating regions the dispersion
relations of the effective waveguide modes lie close to the light
line, we can expect a complicated response because the resonances
associated with the anomalies, considered independently, lie close
to each other. This is considered in some examples presented in section
\ref{sec:Coupled-equations}.

\section{Coupled wave vector equations\label{sec:Coupled-equations}}

We now turn to the solution for the fields in the presence of a grating
$\boldsymbol{\chi}\left(\zeta\right)$ of the form (\ref{eq:Chidecomp}),
where since $\boldsymbol{\chi}_{v}\left(\zeta\right)$ is taken as
periodic with period $a$, we can expand it in a Fourier series
\begin{align*}
\boldsymbol{\chi}_{v}\left(\zeta\right) & =\sum_{m}\boldsymbol{\chi}_{v[m]}e^{im\boldsymbol{\mathbf{K}\cdot\mathbf{R}}},
\end{align*}
where $m$ ranges over the integers and $\mathbf{K}$ is given by
(\ref{eq:Kdef}); here 
\begin{align*}
\boldsymbol{\chi}_{v[m]}= & \frac{1}{a}\int_{-a/2}^{a/2}e^{-imK\zeta}\boldsymbol{\chi}_{v}\left(\zeta\right)d\zeta,
\end{align*}
with $K=\left|\mathbf{K}\right|$. Note that by virtue of the definition
(\ref{eq:chiv_def}) of $\boldsymbol{\chi}_{v}\left(\zeta\right)$
we have $\boldsymbol{\chi}_{v[0]}=0$. Since $\mathbf{P}_{\nu}(\mathbf{R})$
is the response (\ref{eq:Pvdef}) to $\mathbf{E}_{dr}(\mathbf{R})$
due to $\boldsymbol{\chi}_{v}\left(\zeta\right)$, we seek a solution
for our fields of the form 
\begin{align*}
\mathbf{F}\left(\boldsymbol{\kappa}\right)= & \left(2\pi\right)^{2}\sum_{m}\delta\left(\boldsymbol{\kappa}-\boldsymbol{\kappa}_{in}-m\mathbf{K}\right)\mathcal{F}\left(\boldsymbol{\kappa}_{in}+m\mathbf{K}\right),
\end{align*}
where 
\begin{align}
\mathbf{F}\left(\mathbf{R}\right)= & \sum_{m}\mathcal{F}\left(\boldsymbol{\kappa}_{m}\right)e^{i(\boldsymbol{\kappa}_{in}+m\mathbf{K})\cdot\mathbf{R}},\label{eq:mexpansion}
\end{align}
and here and henceforth we put 
\begin{align*}
\boldsymbol{\kappa}_{m}\equiv & \boldsymbol{\kappa}_{in}+m\mathbf{K}.
\end{align*}
Here\textbf{ $\boldsymbol{\kappa}_{in}$} characterizes the incident
field, but we actually allow the incident field $\mathbf{E}_{in}(\mathbf{R})$
to be of the general form (\ref{eq:mexpansion}), with $\mathcal{E}_{in}\left(\boldsymbol{\kappa}_{m}\right)$
nonzero for $m\neq0$; in later sections we will consider a grating
above a substrate, and terms with $m\neq0$ will arise from reflection
of scattered light off the substrate. Using the expansion (\ref{eq:mexpansion})
in (\ref{eq:Pvdef}) we have 
\begin{align}
\mathcal{P}_{\nu}\left(\boldsymbol{\kappa}_{m}\right)= & \epsilon_{0}\sum_{m'}\boldsymbol{\chi}_{v[m-m']}\cdot\mathcal{E}_{mod}(\boldsymbol{\kappa}_{m'}),\label{eq:chiGrating}
\end{align}
for example; equations for the Fourier components of other quantities
will be given below. The set of these equations can be organized as
matrix equations in many ways; below we present one approach that
is both useful for calculations, and allows for an easy proof of energy
conservation even when the number of Fourier components is truncated.

\subsection{S-matrix equations\label{subsec:S-matrix-equations}}

To complete a calculation we approximate sums over $m$ by a restriction
to $\left|m\right|\leq N$, where the threshold integer $N$ includes
at least all diffracted, propagating orders. For each field $\mathbf{F}(\mathbf{R})$
we then introduce $\bar{\mathcal{F}}$, a column of columns 
\begin{align}
\bar{\mathcal{F}}= & \left[\begin{array}{c}
\bar{\mathcal{F}}(\boldsymbol{\kappa}_{N})\\
\bar{\mathcal{F}}(\boldsymbol{\kappa}_{(N-1)})\\
\vdots\\
\bar{\mathcal{F}}(\boldsymbol{\kappa}_{(-N)})
\end{array}\right],\label{eq:Ffull}
\end{align}
where each $\bar{\mathcal{F}}\left(\boldsymbol{\kappa}_{m}\right)$
is a column with the three Cartesian components of $\mathcal{F}\left(\boldsymbol{\kappa}_{m}\right)$,
\begin{align}
\bar{\mathcal{F}}\left(\boldsymbol{\kappa}_{m}\right)= & \left[\begin{array}{c}
\hat{\boldsymbol{x}}\cdot\mathcal{F}\left(\boldsymbol{\kappa}_{m}\right)\\
\hat{\boldsymbol{y}}\cdot\mathcal{F}\left(\boldsymbol{\kappa}_{m}\right)\\
\hat{\boldsymbol{z}}\cdot\mathcal{F}\left(\boldsymbol{\kappa}_{m}\right)
\end{array}\right],\label{eq:Fsub}
\end{align}
and so the full column $\bar{\mathcal{F}}$ has $3(2N+1)$ elements.
For the tensors we introduce $(2N+1)\times(2N+1)$ matrices with elements
that are themselves $3\times3$ matrices; thus in each of these there
are $3(2N+1)\times3(2N+1)$ elements in all. We put 
\begin{align*}
\bar{g}= & \left[\begin{array}{cccc}
\bar{g}_{NN} & \bar{0} & \cdots & \bar{0}\\
\bar{0} & \bar{g}{}_{(N-1)(N-1)} & \cdots & \bar{0}\\
\vdots & \vdots & \ddots & \vdots\\
\bar{0} & \bar{0} & \cdots & \bar{g}_{(-N)(-N)}
\end{array}\right],
\end{align*}
a block diagonal matrix where $\bar{0}$ indicates a $3\times3$ matrix
of zeros, and the $3\times3$ matrices $\bar{g}_{mm}$ are given by
\begin{align*}
\bar{g}_{mm}= & \left[\begin{array}{ccc}
\left(\hat{\boldsymbol{x}}\cdot\mathbf{g}\left(\boldsymbol{\kappa}_{m}\right)\cdot\hat{\boldsymbol{x}}\right) & \left(\hat{\boldsymbol{x}}\cdot\mathbf{g}\left(\boldsymbol{\kappa}_{m}\right)\cdot\hat{\boldsymbol{y}}\right) & \left(\hat{\boldsymbol{x}}\cdot\mathbf{g}\left(\boldsymbol{\kappa}_{m}\right)\cdot\hat{\boldsymbol{z}}\right)\\
\left(\hat{\boldsymbol{y}}\cdot\mathbf{g}\left(\boldsymbol{\kappa}_{m}\right)\cdot\hat{\boldsymbol{x}}\right) & \left(\hat{\boldsymbol{y}}\cdot\mathbf{g}\left(\boldsymbol{\kappa}_{m}\right)\cdot\hat{\boldsymbol{y}}\right) & \left(\hat{\boldsymbol{y}}\cdot\mathbf{g}\left(\boldsymbol{\kappa}_{m}\right)\cdot\hat{\boldsymbol{z}}\right)\\
\left(\hat{\boldsymbol{z}}\cdot\mathbf{g}\left(\boldsymbol{\kappa}_{m}\right)\cdot\hat{\boldsymbol{x}}\right) & \left(\hat{\boldsymbol{z}}\cdot\mathbf{g}\left(\boldsymbol{\kappa}_{m}\right)\cdot\hat{\boldsymbol{y}}\right) & \left(\hat{\boldsymbol{z}}\cdot\mathbf{g}\left(\boldsymbol{\kappa}_{m}\right)\cdot\hat{\boldsymbol{z}}\right)
\end{array}\right].
\end{align*}

\begin{figure}
\centering

\includegraphics[width=0.5\columnwidth]{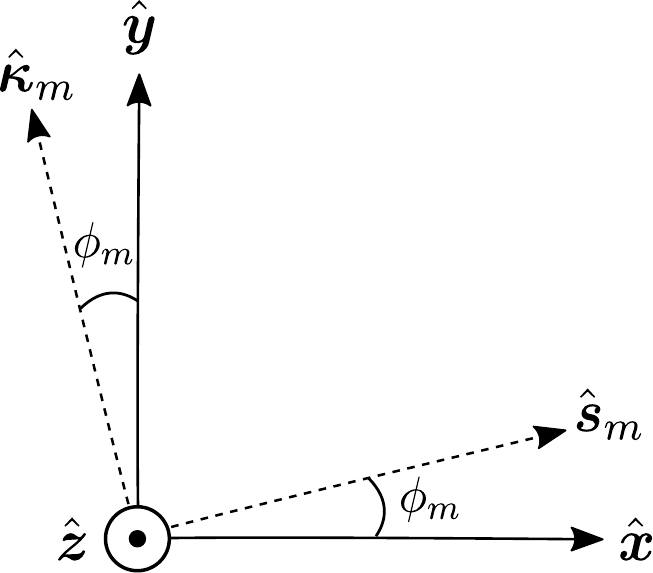}

\caption{Plane containing the set of axes $\left(\hat{\boldsymbol{x}},\hat{\boldsymbol{y}}\right)$
and basis vectors $\left(\hat{\boldsymbol{s}}_{m},\hat{\boldsymbol{\kappa}}_{m}\right)$. }

\label{fig:axes}
\end{figure}

For example, let the associated polarization vectors associated with
$\boldsymbol{\kappa}_{m}$ be $\hat{\boldsymbol{s}}_{m}$ and $\hat{\boldsymbol{p}}_{m\pm}$,
such that 
\begin{align}
\hat{\boldsymbol{s}}_{m} & =\hat{\boldsymbol{\kappa}}_{m}\times\hat{\boldsymbol{z}},\label{eq:unit_vectors}\\
\hat{\boldsymbol{p}}_{1\pm,m} & =\frac{\kappa_{m}\hat{\boldsymbol{z}}\mp w_{1}\left(\boldsymbol{\kappa}_{m}\right)\hat{\boldsymbol{\kappa}}_{m}}{\tilde{\omega}n_{1}},\nonumber 
\end{align}
with $\hat{\boldsymbol{\kappa}}_{m}=\boldsymbol{\kappa}_{m}/\left|\boldsymbol{\kappa}_{m}\right|$
and $w_{1}\left(\boldsymbol{\kappa}_{m}\right)=\sqrt{\tilde{\omega}^{2}\varepsilon_{1}-\kappa_{m}^{2}}$
(compare (\ref{eq:unit_def})). If we then let $\phi_{m}$ indicate
the rotation in the $xy$ plane between the sets of unit orthogonal
vectors $(\hat{\boldsymbol{x}},\hat{\boldsymbol{y}})$ and $(\mathbf{\hat{s}}_{m},\hat{\boldsymbol{\kappa}}_{m})$
(see Fig. \ref{fig:axes}), we have 
\begin{align*}
\bar{g}{}_{mm}= & \frac{i\tilde{\omega}^{2}D}{2\epsilon_{0}w_{1}\left(\boldsymbol{\kappa}_{m}\right)}\left[\begin{array}{ccc}
1-\frac{\kappa_{m}}{\tilde{\omega}^{2}\varepsilon_{1}}\sin^{2}\phi_{m} & \frac{\kappa_{m}}{2\tilde{\omega}^{2}\varepsilon_{1}}\sin2\phi_{m} & 0\\
\frac{\kappa_{m}}{2\tilde{\omega}^{2}\varepsilon_{1}}\sin2\phi_{m} & 1-\frac{\kappa_{m}}{\tilde{\omega}^{2}\varepsilon_{1}}\cos^{2}\phi_{m} & 0\\
0 & 0 & \frac{\kappa_{m}}{\tilde{\omega}^{2}\varepsilon_{1}}
\end{array}\right].
\end{align*}
Block diagonal matrices $\bar{G}^{\pm}$ and $\bar{\chi}_{o}$ are
defined similarly, where the blocks $\bar{\chi}_{o;mm}$ of $\bar{\chi}_{o}$
are all identical, $\bar{\chi}_{o;mm}=diag(\left\langle \chi_{mod}^{xx}\right\rangle ,\left\langle \chi_{mod}^{yy}\right\rangle ,\left\langle \chi_{mod}^{zz}\right\rangle )$.
The matrix of matrices $\bar{\chi}_{v}$ representing the grating
is not block diagonal, but is given by 
\begin{align}
\bar{\chi}_{v} & =\left[\begin{array}{cccc}
\bar{0} & \bar{\chi}_{v;N(N-1)} & \cdots & \bar{\chi}_{\nu;N(-N)}\\
\bar{\chi}_{v;(N-1)(N)} & \bar{0} & \cdots & \bar{\chi}_{v;(N-1)(-N)}\\
\vdots & \vdots & \ddots & \vdots\\
\bar{\chi}_{v;(-N)(N)} & \bar{\chi}_{v;(-N)(N-1)} & \cdots & \bar{0}
\end{array}\right],\label{eq:chi_v_bar}
\end{align}
where 
\[
\bar{\chi}_{v;mm'}=\begin{bmatrix}\hat{\boldsymbol{x}}\cdot\boldsymbol{\chi}_{v[m-m']}\cdot\hat{\boldsymbol{x}} & 0 & 0\\
0 & \hat{\boldsymbol{y}}\cdot\boldsymbol{\chi}_{v[m-m']}\cdot\hat{\boldsymbol{y}} & 0\\
0 & 0 & \hat{\boldsymbol{z}}\cdot\boldsymbol{\chi}_{v[m-m']}\cdot\hat{\boldsymbol{z}}
\end{bmatrix},
\]
and the diagonal elements of $\bar{\chi}_{v}$ vanish because $\boldsymbol{\chi}_{v[0]}=0$.
In this notation the equations (\ref{eq:chiGrating}) for the $\mathcal{P}_{\nu}\left(\boldsymbol{\kappa}_{m}\right)$
can be written in full matrix form as 
\begin{align*}
\bar{\mathcal{P}}_{v}= & \epsilon_{0}\bar{\chi}_{v}\bar{\mathcal{E}}_{mod},
\end{align*}
and combining this with the matrix form of (\ref{eq:soe}) we find
\begin{align*}
\bar{\mathcal{P}} & =\epsilon_{0}(\bar{\chi}_{o}+\bar{\chi}_{v})\bar{\mathcal{E}}_{mod},\\
\bar{\mathcal{E}}_{mod} & =\bar{\mathcal{E}}_{in}+\bar{g}\bar{\mathcal{P}},
\end{align*}
with a formal solution 
\begin{align*}
\bar{\mathcal{P}}= & \epsilon_{0}\left(\bar{\chi}_{o}+\bar{\chi}_{v}\right)\left[\bar{1}_{3}-\epsilon_{0}\bar{g}(\bar{\chi}_{o}+\bar{\chi}_{v})\right]^{-1}\bar{\mathcal{E}}_{in},
\end{align*}
where $\bar{1}_{j}$ denotes the $j(2N+1)\times j(2N+1)$ unit matrix,
with $j$ an integer. Introducing columns $\bar{\mathcal{E}}_{sc}^{\pm}$
to describe the scattered fields, from (\ref{eq:Escat_def}) we then
have 
\begin{align*}
\bar{\mathcal{E}}_{sc}^{\pm}= & \epsilon_{0}\bar{G}^{\pm}\left(\bar{\chi}_{o}+\bar{\chi}_{v}\right)\left[\bar{1}_{3}-\epsilon_{0}\bar{g}(\bar{\chi}_{o}+\bar{\chi}_{v})\right]^{-1}\bar{\mathcal{E}}_{in}.
\end{align*}
Separating out the upward propagating (or evanescent) contributions
of the incident field from the corresponding downward propagations
(see (\ref{eq:Ein_def})), we have $\bar{\mathcal{E}}_{in}=\bar{\mathcal{E}}_{in}^{+}+\bar{\mathcal{E}}_{in}^{-}$,
and introducing columns for the full outward propagating (or evanescent)
fields for $z>D/2$ and $z<-D/2$ (see (\ref{eq:Eout_def})), with
new columns defined as indicated we can write 
\begin{align}
\bar{\mathcal{E}}_{out}^{+} & =\epsilon_{0}\bar{G}^{+}\left(\bar{\chi}_{o}+\bar{\chi}_{v}\right)\left[\bar{1}_{3}-\epsilon_{0}\bar{g}(\bar{\chi}_{o}+\bar{\chi}_{v})\right]^{-1}\left(\bar{\mathcal{E}}_{in}^{+}+\bar{\mathcal{E}}_{in}^{-}\right)+\bar{\mathcal{E}}_{in}^{+},\label{eq:basic_scattering}\\
\bar{\mathcal{E}}_{out}^{-} & =\epsilon_{0}\bar{G}^{-}\left(\bar{\chi}_{o}+\bar{\chi}_{v}\right)\left[\bar{1}_{3}-\epsilon_{0}\bar{g}(\bar{\chi}_{o}+\bar{\chi}_{v})\right]^{-1}\left(\bar{\mathcal{E}}_{in}^{+}+\bar{\mathcal{E}}_{in}^{-}\right)+\bar{\mathcal{E}}_{in}^{-}.\nonumber 
\end{align}

The sub-columns of $\bar{\mathcal{E}}_{in}^{\pm}$, $\bar{\mathcal{E}}_{in}^{\pm}\left(\boldsymbol{\kappa}_{m}\right)$,
contain the three Cartesian components of $\mathcal{E}_{in}^{\pm}\left(\boldsymbol{\kappa}_{m}\right)$
(recall (\ref{eq:Fsub})). However, these are not independent, since
for any $\boldsymbol{\kappa}_{m}$ there are only $s-$ and $p-$polarized
components, 
\begin{align*}
\mathcal{E}_{in}^{\pm}\left(\boldsymbol{\kappa}_{m}\right)= & \hat{\boldsymbol{s}}_{m}\mathbb{E}_{in;s}^{\pm}\left(\boldsymbol{\kappa}_{m}\right)+\hat{\boldsymbol{p}}_{1\pm,m}\mathbb{E}_{in;p}^{\pm}\left(\boldsymbol{\kappa}_{m}\right),
\end{align*}
with two independent amplitudes $\mathbb{E}_{in;s}^{\pm}\left(\boldsymbol{\kappa}_{m}\right)$
and $\mathbb{E}_{in;p}^{\pm}\left(\boldsymbol{\kappa}_{m}\right)$.
As such we can write 
\begin{align*}
\bar{\mathcal{E}}_{in}^{\pm}\left(\boldsymbol{\kappa}_{m}\right) & =\left[\begin{array}{c}
\hat{\boldsymbol{x}}\cdot\mathcal{E}_{in}^{\pm}\left(\boldsymbol{\kappa}_{m}\right)\\
\hat{\boldsymbol{y}}\cdot\mathcal{E}_{in}^{\pm}\left(\boldsymbol{\kappa}_{m}\right)\\
\hat{\boldsymbol{z}}\cdot\mathcal{E}_{in}^{\pm}\left(\boldsymbol{\kappa}_{m}\right)
\end{array}\right],\\
 & =\bar{\sigma}_{in}^{\pm}\left(\boldsymbol{\kappa}_{m}\right)\mathbb{\bar{\mathbb{E}}}_{in}^{\pm}\left(\boldsymbol{\kappa}_{m}\right),
\end{align*}
where 
\begin{align}
\bar{\sigma}_{in}^{\pm}\left(\boldsymbol{\kappa}_{m}\right)= & \left[\begin{array}{cc}
\left(\hat{\boldsymbol{x}}\cdot\hat{\boldsymbol{s}}_{m}\right) & \left(\hat{\boldsymbol{x}}\cdot\hat{\boldsymbol{p}}_{1\pm,m}\right)\\
\left(\hat{\boldsymbol{y}}\cdot\hat{\boldsymbol{s}}_{m}\right) & \left(\hat{\boldsymbol{y}}\cdot\hat{\boldsymbol{p}}_{1\pm,m}\right)\\
\left(\hat{\boldsymbol{z}}\cdot\hat{\boldsymbol{s}}_{m}\right) & \left(\hat{\boldsymbol{z}}\cdot\hat{\boldsymbol{p}}_{1\pm,m}\right)
\end{array}\right]\label{eq:sigma_in_terms}
\end{align}
is a $3\times2$ matrix for each of the $(+)$ and $(-)$ examples,
and 
\begin{align}
\mathbb{\bar{\mathbb{E}}}_{in}^{\pm}\left(\boldsymbol{\kappa}_{m}\right)= & \left[\begin{array}{c}
\mathbb{E}_{in;s}^{\pm}\left(\boldsymbol{\kappa}_{m}\right)\\
\mathbb{E}_{in;p}^{\pm}\left(\boldsymbol{\kappa}_{m}\right)
\end{array}\right]\label{eq:Ein_bit}
\end{align}
is a column of two elements. Constructing the full column for all
$\boldsymbol{\kappa}_{m}$ components of $\bar{\mathcal{E}}_{in}^{\pm}$
we have (recall (\ref{eq:Ffull}))
\begin{align}
\bar{\mathcal{E}}_{in}^{\pm}= & \bar{\sigma}_{in}^{\pm}\bar{\mathbb{E}}_{in}^{\pm},\label{eq:in_relate}
\end{align}
where 
\begin{align}
\bar{\mathbb{E}}_{in}^{\pm}= & \left[\begin{array}{c}
\mathbb{\bar{\mathbb{E}}}_{in}^{\pm}(\boldsymbol{\kappa}_{N})\\
\mathbb{\mathbb{\bar{\mathbb{E}}}}_{in}^{\pm}(\boldsymbol{\kappa}_{(N-1)})\\
\vdots\\
\mathbb{\bar{\mathbb{E}}}_{in}^{\pm}(\boldsymbol{\kappa}_{(-N)})
\end{array}\right],\label{eq:Ein_full}
\end{align}
which for each of the $(+)$ and $(-)$ examples is a column with
$2(2N+1)$ elements, once all the $\mathbb{\bar{\mathbb{E}}}_{in}^{\pm}\left(\boldsymbol{\kappa}_{m}\right)$
are written out. Further, 
\begin{align}
\bar{\sigma}_{in}^{\pm}= & \left[\begin{array}{cccc}
\bar{\sigma}_{in}^{\pm}(\boldsymbol{\kappa}_{N}) & \bar{0} & \cdots & \bar{0}\\
\bar{0} & \bar{\sigma}_{in}^{\pm}(\boldsymbol{\kappa}_{(N-1)}) & \cdots & \bar{0}\\
\vdots & \vdots & \ddots & \vdots\\
\bar{0} & \bar{0} & \cdots & \bar{\sigma}_{in}^{\pm}(\boldsymbol{\kappa}_{(-N)})
\end{array}\right],\label{eq:sigma_in_big}
\end{align}
which for each of the examples is a $3(2N+1)\times2(2N+1)$ matrix,
once all the elements of the $\bar{\sigma}_{in}^{\pm}\left(\boldsymbol{\kappa}_{m}\right)$
are written out; here $\bar{0}$ are $3\times2$ matrices with all
elements vanishing. 

Similarly, for each $\bar{\mathcal{E}}_{out}^{\pm}$$\left(\boldsymbol{\kappa}_{m}\right)$
in $\bar{\mathcal{E}}_{out}^{\pm}$ there will be only $s-$ and $p-$polarized
components, 
\begin{align*}
\mathcal{\mathcal{E}}_{out}^{\pm}\left(\boldsymbol{\kappa}_{m}\right)= & \hat{\boldsymbol{s}}_{m}\mathbb{\mathbb{E}}_{out;s}^{\pm}\left(\boldsymbol{\kappa}_{m}\right)+\hat{\boldsymbol{p}}_{1\pm,m}\mathbb{\mathbb{E}}_{out;p}^{\pm}\left(\boldsymbol{\kappa}_{m}\right),
\end{align*}
which we can immediately see will be identified by the ${\bf G}^{\pm}\left(\boldsymbol{\kappa}_{m}\right)$
(see (\ref{eq:Green})) that appear in $\bar{G}^{\pm}.$ Nonetheless,
we can formally extract those amplitudes $\mathcal{\mathbb{E}}_{out;s,p}^{\pm}\left(\boldsymbol{\kappa}_{m}\right)$
by writing 
\begin{align}
\bar{\mathbb{E}}_{out}^{\pm}= & \bar{\sigma}_{out}^{\pm}\bar{\mathcal{E}}_{out}^{\pm},\label{eq:outrelate}
\end{align}
where 
\begin{align*}
\bar{\mathbb{E}}_{out}^{\pm}= & \left[\begin{array}{c}
\bar{\mathbb{E}}_{out}^{\pm}(\boldsymbol{\kappa}_{N})\\
\bar{\mathbb{E}}_{out}^{\pm}(\boldsymbol{\kappa}_{(N-1)})\\
\vdots\\
\bar{\mathbb{E}}_{out}^{\pm}(\boldsymbol{\kappa}_{(-N)})
\end{array}\right],
\end{align*}
is a $2(2N+1)$ column for each example $(\pm$) once the elements
of 
\begin{align*}
\bar{\mathbb{E}}_{out}^{\pm}\left(\boldsymbol{\kappa}_{m}\right)= & \left[\begin{array}{c}
\mathbb{E}_{out;s}^{\pm}\left(\boldsymbol{\kappa}_{m}\right)\\
\mathbb{E}_{out;p}^{\pm}\left(\boldsymbol{\kappa}_{m}\right)
\end{array}\right]
\end{align*}
are written out, and 
\begin{align}
\bar{\sigma}_{out}^{\pm}= & \left[\begin{array}{cccc}
\bar{\sigma}_{out}^{\pm}(\boldsymbol{\kappa}_{N}) & \bar{0} & \cdots & \bar{0}\\
\bar{0} & \bar{\sigma}_{out}^{\pm}(\boldsymbol{\kappa}_{(N-1)}) & \cdots & \bar{0}\\
\vdots & \vdots & \ddots & \vdots\\
\bar{0} & \bar{0} & \cdots & \bar{\sigma}_{out}^{\pm}(\boldsymbol{\kappa}_{(-N)})
\end{array}\right],\label{eq:sigma_out_big}
\end{align}
where the $\bar{0}$ denote $2\times3$ matrices with all their elements
vanishing, and 
\begin{align}
\bar{\sigma}_{out}^{\pm}\left(\boldsymbol{\kappa}_{m}\right)= & \left[\begin{array}{ccc}
\left(\hat{\boldsymbol{x}}\cdot\hat{\boldsymbol{s}}_{m}\right) & \left(\hat{\boldsymbol{y}}\cdot\hat{\boldsymbol{s}}_{m}\right) & \left(\hat{\boldsymbol{z}}\cdot\hat{\boldsymbol{s}}_{m}\right)\\
\left(\hat{\boldsymbol{x}}\cdot\hat{\boldsymbol{p}}_{1\pm,m}\right) & \left(\hat{\boldsymbol{y}}\cdot\hat{\boldsymbol{p}}_{1\pm,m}\right) & \left(\hat{\boldsymbol{z}}\cdot\hat{\boldsymbol{p}}_{1\pm,m}\right)
\end{array}\right].\label{eq:sigma_out_terms}
\end{align}
Using (\ref{eq:in_relate},\ref{eq:outrelate}) in (\ref{eq:basic_scattering}),
we can write 
\begin{align*}
\bar{\mathbb{E}}_{out}^{+} & =\bar{\mathbb{T}}_{g}\mathbb{\overline{E}}_{in}^{+}+\bar{\mathbb{R}}_{g}\mathbb{\overline{E}}_{in}^{-},\\
\bar{\mathbb{E}}_{out}^{-} & =\bar{\mathbb{R}}_{g}\bar{\mathbb{E}}_{in}^{+}+\bar{\mathbb{T}}_{g}\bar{\mathbb{E}}_{in}^{-},
\end{align*}
where $\bar{\mathbb{R}}_{g}$ and $\bar{\mathbb{T}}_{g}$ are $2(2N+1)\times2(2N+1)$
matrices, 
\begin{equation}
\begin{array}{rl}
\bar{\mathbb{T}}_{g}= & \bar{\sigma}_{out}^{\pm}\left(\bar{1}_{3}+\epsilon_{0}\bar{G}^{\pm}\left(\bar{\chi}_{o}+\bar{\chi}_{v}\right)\left[\bar{1}_{3}-\epsilon_{0}\bar{g}\left(\bar{\chi}_{o}+\bar{\chi}_{v}\right)\right]^{-1}\right)\bar{\sigma}_{in}^{\pm},\\
\bar{\mathbb{R}}_{g}= & \bar{\sigma}_{out}^{\pm}\epsilon_{0}\bar{G}^{\pm}\left(\bar{\chi}_{o}+\bar{\chi}_{v}\right)\left[\bar{1}_{3}-\epsilon_{0}\bar{g}\left(\bar{\chi}_{o}+\bar{\chi}_{v}\right)\right]^{-1}\bar{\sigma}_{in}^{\mp}.
\end{array}\label{eq:scatmat}
\end{equation}
Since we consider the same dielectric above and below the grating,
the transmission and reflection properties are the same whether light
is incident from above or below; thus the expressions (\ref{eq:scatmat})
are the same whether the $+$ or $-$ matrices on the right-hand-side
of the equations are used in their evaluation.

Finally, combining the two columns $\mathbb{\bar{\mathbb{E}}}_{out}^{+}$
and $\bar{\mathbb{E}}_{out}^{-}$, each with $2(2N+1)$ elements,
to form one column with $4(2N+1)$ elements, and likewise for $\bar{\mathbb{E}}_{in}^{+}$
and $\bar{\mathbb{E}}_{in}^{-}$, we can form a $4(2N+1)\times4(2N+1)$
scattering matrix $\mathcal{\mathscr{S}}$, 
\begin{equation}
\mathscr{S}\equiv\left[\begin{array}{cc}
\bar{\mathbb{T}}_{g} & \bar{\mathbb{R}}_{g}\\
\bar{\mathbb{R}}_{g} & \bar{\mathbb{T}}_{g}
\end{array}\right],\label{eq:S}
\end{equation}
so that 
\begin{align}
\left[\begin{array}{c}
\bar{\mathbb{E}}_{out}^{+}\\
\bar{\mathbb{E}}_{out}^{-}
\end{array}\right]= & \mathscr{S}\left[\begin{array}{c}
\bar{\mathbb{E}}_{in}^{+}\\
\bar{\mathbb{E}}_{in}^{-}
\end{array}\right].\label{eq:full_scattering}
\end{align}
With the equations in this form, a proof of energy conservation is
possible, and is presented in Appendix \ref{sec:conserve}. That proof,
and the equations from which it was derived, hold for any orientation
of the grating direction $\hat{\boldsymbol{e}}$ in the $xy$ plane
and any plane of incidence.

\subsection{A simple configuration\label{subsec:simple-config}}

\begin{figure}
\centering

\includegraphics[width=0.8\columnwidth]{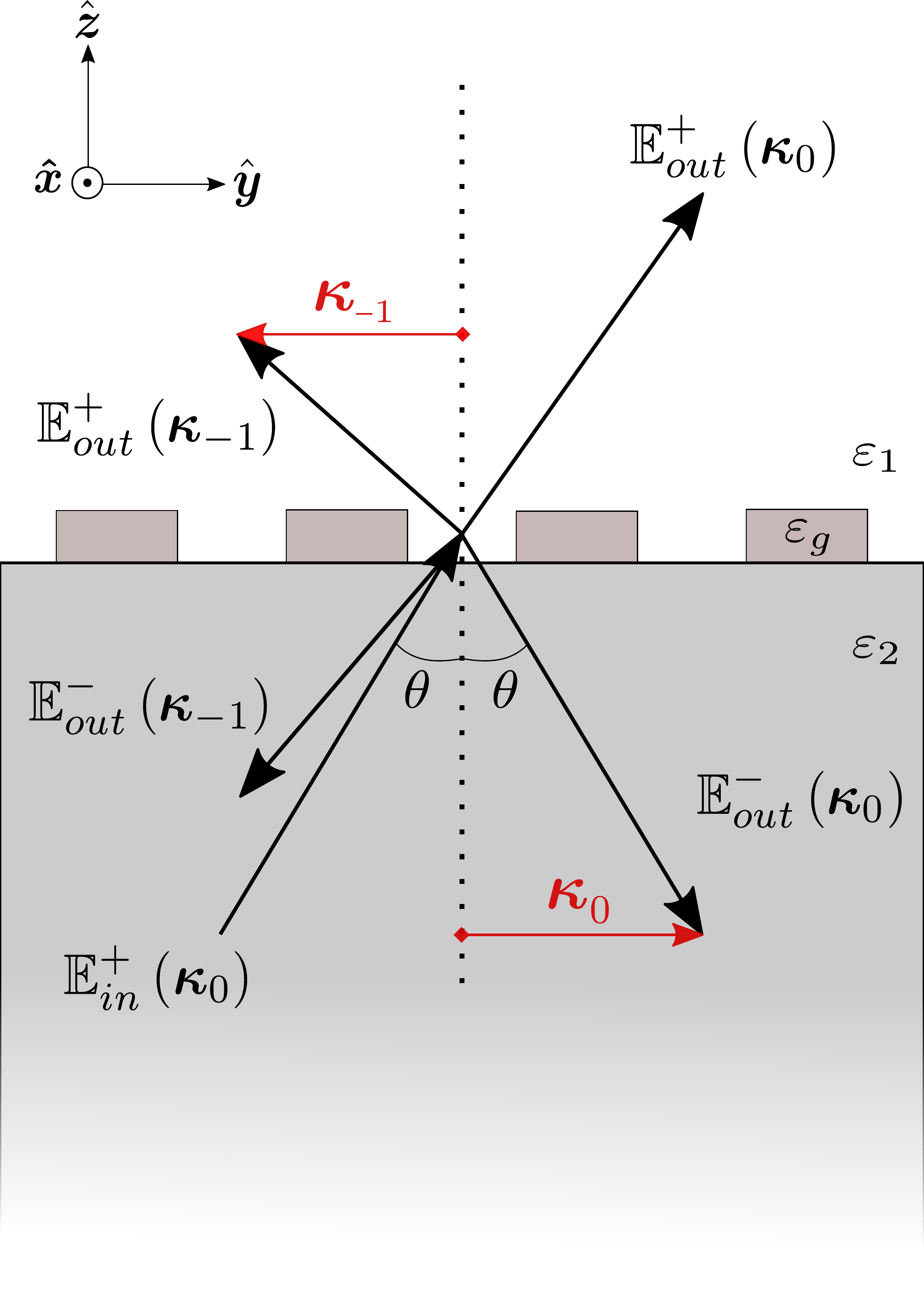}

\caption{(Color online) A simple 1D grating configuration with the grating
oriented such that $\hat{\boldsymbol{e}}=\hat{\boldsymbol{y}}$. Incident,
reflected, transmitted, and diffracted rays are shown by black (thick)
lines, and labeled by the notation used to indicate their field amplitudes;
the projection of the wave vectors on the $xy$ plane are shown by
red (thin) lines.}

\label{fig:simple_cfg}
\end{figure}

In this subsection we simplify the equations above for a common scenario
of interest: We take the grating susceptibility to be uniaxial, $\chi_{add}^{xx}\left(\zeta\right)=\chi_{add}^{yy}\left(\zeta\right)$
(recall (\ref{eq:Chidecomp},\ref{eq:chi_add})), choose $\hat{\boldsymbol{e}}=\hat{\boldsymbol{y}},$
and assume the plane of incidence contains $\hat{\boldsymbol{z}}$
and $\hat{\boldsymbol{e}}=\hat{\boldsymbol{y}}$, as illustrated in
Fig. \ref{fig:simple_cfg}; for the isolated grating treated above
and in this section, we have $\varepsilon_{2}=\varepsilon_{1}$. The
wave vectors $\boldsymbol{\kappa}_{m}$ that are relevant here are
then either in the $\hat{\boldsymbol{y}}$ or $-\hat{\boldsymbol{y}}$
direction, so \textbf{$\hat{\boldsymbol{\kappa}}_{m}\cdot\hat{\boldsymbol{y}}=sign(\hat{\boldsymbol{\kappa}}_{m}\cdot\hat{\boldsymbol{y}})=\hat{\boldsymbol{s}}_{m}\cdot\hat{\boldsymbol{x}}$}
; the form of the expressions (\ref{eq:sigma_in_terms}) for the $\bar{\sigma}_{in}^{\pm}\left(\boldsymbol{\kappa}_{m}\right)$
simplifies to 
\begin{align*}
\bar{\sigma}_{in}^{\pm}\left(\boldsymbol{\kappa}_{m}\right)= & \left[\begin{array}{cc}
(\hat{\boldsymbol{\kappa}}_{m}\cdot\hat{\boldsymbol{y}}) & 0\\
0 & \mp\frac{w_{1}\left(\kappa_{m}\right)}{\tilde{\omega}n_{1}}(\hat{\boldsymbol{\kappa}}_{m}\cdot\hat{\boldsymbol{y}})\\
0 & \frac{\kappa_{m}}{\tilde{\omega}n_{1}}
\end{array}\right],
\end{align*}
and similarly for the form of the expressions (\ref{eq:sigma_out_terms})
for $\bar{\sigma}_{out}^{\pm}\left(\boldsymbol{\kappa}_{m}\right)$,
which are the transpose of the $\bar{\sigma}_{in}^{\pm}\left(\boldsymbol{\kappa}_{m}\right)$.
When these are assembled into $\bar{\sigma}_{in}^{\pm}$ and $\bar{\sigma}_{out}^{\pm}$
in (\ref{eq:sigma_in_big},\ref{eq:sigma_out_big}) and the results
used in (\ref{eq:scatmat}) for $\bar{\mathbb{R}}_{g}$and $\bar{\mathbb{T}}_{g}$,
we find that because of the high symmetry of the problem each of these
$2(2N+1)\times2(2N+1)$ matrices can be reorganized into two $(2N+1)\times(2N+1)$
matrices, one relevant for $s$-polarized light and one for $p$-polarized
light. For each polarization the relevant matrices can then be combined
into a $2(2N+1)\times2(2N+1)$ scattering matrix, and in place of
(\ref{eq:full_scattering}) we have two sets of equations, 
\begin{align}
\left[\begin{array}{c}
\bar{\mathbb{E}}_{out,\alpha}^{+}\\
\bar{\mathbb{E}}_{out,\alpha}^{-}
\end{array}\right]= & \mathscr{S}_{\alpha}\left[\begin{array}{c}
\bar{\mathbb{E}}_{in,\alpha}^{+}\\
\bar{\mathbb{E}}_{in,\alpha}^{-}
\end{array}\right],\label{eq:scat_simple}
\end{align}
where $\alpha=s,p$, each $\bar{\mathbb{E}}_{in,\alpha}^{+}$ is a
$(2N+1)$ element column, 
\begin{align}
\bar{\mathbb{E}}_{in,\alpha}^{-}= & \left[\begin{array}{c}
\mathbb{E}_{in,\alpha}^{-}(\boldsymbol{\kappa}_{N})\\
\mathbb{E}_{in,\alpha}^{-}(\boldsymbol{\kappa}_{N-1})\\
\vdots\\
\mathbb{E}_{in,\alpha}^{-}(\boldsymbol{\kappa}_{-N})
\end{array}\right],\label{eq:field_simple}
\end{align}
(compare (\ref{eq:Ein_bit},\ref{eq:Ein_full})), and likewise for
$\bar{\mathbb{E}}_{in,\alpha}^{+}$ and $\bar{\mathbb{E}}_{out,\alpha}^{\pm}$,
and where 
\begin{align*}
\mathscr{S}_{\alpha}\equiv & \left[\begin{array}{cc}
\mathbb{T}_{g,\alpha} & \mathbb{R}_{g,\alpha}\\
\mathbb{R}_{g,\alpha} & \mathbb{T}_{g,\alpha}
\end{array}\right],
\end{align*}
with 
\begin{align}
\mathbb{T}_{g,s} & =\bar{\beta}\left[\bar{1}_{1}-\frac{i\tilde{\omega}^{2}D}{2}\bar{w}_{1}^{-1}\bar{\chi}^{\parallel}\right]^{-1}\bar{\beta},\label{eq:TRs}\\
\mathbb{R}_{g,s} & =\mathbb{T}_{g,s}-\bar{1}_{1},\nonumber 
\end{align}
and 
\begin{align}
\mathbb{T}_{g,p} & =\left[\bar{1}_{1}-\frac{iD}{2\varepsilon_{1}}\bar{\kappa}\bar{\chi}^{\perp}\bar{\kappa}\right]^{-1}+\bar{\beta}\left[\bar{1}_{1}-\frac{iD}{2\varepsilon_{1}}\bar{\chi}^{\parallel}\bar{w}_{1}\right]^{-1}\bar{\beta}-\bar{1}_{1},\label{eq:TRp}\\
\mathbb{R}_{g,p} & =\left[\bar{1}_{1}-\frac{iD}{2\varepsilon_{1}}\bar{\kappa}\bar{\chi}^{\perp}\bar{\kappa}\right]^{-1}-\bar{\beta}\left[\bar{1}_{1}-\frac{iD}{2\varepsilon_{1}}\bar{\chi}^{\parallel}\bar{w}_{1}\right]^{-1}\bar{\beta}.\nonumber 
\end{align}
Here each of $\mathbb{T}_{g,s}$, $\mathbb{R}_{g,s}$, $\mathbb{T}_{g,p}$,
and $\mathbb{R}_{g,p}$ is a $(2N+1)\times(2N+1)$ matrix. The matrices
$\bar{\beta}$, $\bar{\kappa},$ and $\bar{w}_{1}$ are diagonal matrices
of the same dimension, $\bar{\beta}=diag(\hat{\boldsymbol{\kappa}}_{N}\cdot\hat{\boldsymbol{y}},\hat{\boldsymbol{\kappa}}_{N-1}\cdot\hat{\boldsymbol{y}},...,\hat{\boldsymbol{\kappa}}_{-N}\cdot\hat{\boldsymbol{y}})$,
$\bar{\kappa}=diag(\left|\boldsymbol{\kappa}_{N}\right|,\left|\boldsymbol{\kappa}_{N-1}\right|,...,\left|\boldsymbol{\kappa}_{-N}\right|),$
and $\bar{w}_{1}=diag(w_{1}(\boldsymbol{\kappa}_{N}),w_{1}(\boldsymbol{\kappa}_{N-1})...,w_{1}(\boldsymbol{\kappa}_{-N})),$
with $w_{1}\left(\boldsymbol{\kappa}\right)=\sqrt{\tilde{\omega}^{2}n_{1}^{2}-\kappa^{2}}$.
Finally, $\bar{\chi}^{\parallel}$ and $\bar{\chi}^{\perp}$ are $(2N+1)\times(2N+1)$
matrices with $(mm')$ elements $\bar{\chi}_{mm'}^{\parallel}=\delta_{mm'}\left\langle \chi_{mod}^{xx}\right\rangle +\left(\hat{\boldsymbol{x}}\cdot\boldsymbol{\chi}_{v[m-m']}\cdot\hat{\boldsymbol{x}}\right)=\delta_{mm'}\left\langle \chi_{mod}^{yy}\right\rangle +\left(\hat{\boldsymbol{y}}\cdot\boldsymbol{\chi}_{v[m-m']}\cdot\hat{\boldsymbol{y}}\right)$
and $\bar{\chi}_{mm'}^{\perp}=\delta_{mm'}\left\langle \chi_{mod}^{zz}\right\rangle +\left(\hat{\boldsymbol{z}}\cdot\boldsymbol{\chi}_{v[m-m']}\cdot\hat{\boldsymbol{z}}\right)$.
We note that the relation between $\mathbb{T}_{s}$ and $\mathbb{R}_{s}$
is simple because the reference vectors $\hat{\boldsymbol{s}}_{m}$
for the fields are all the same or differ simply by a minus sign;
while that between $\mathbb{T}_{p}$ and $\mathbb{R}_{p}$ is more
complicated because, even for a particular $\boldsymbol{\kappa}_{m}$,
the $z$ components of $\hat{\boldsymbol{p}}_{1+,m}$ and $\hat{\boldsymbol{p}}_{1-,m}$
are identical, but the $y$ components differ by a sign (see (\ref{eq:unit_vectors})).

\subsection{An example\label{subsec:An-example}}

The expressions (\ref{eq:TRs},\ref{eq:TRp}) for the reflection and
transmission matrices, and indeed the more general expressions (\ref{eq:scatmat}),
can be used to calculate specular reflection and transmission, and
diffraction, for the choice of any number $2N+1$ of wave vectors
$\boldsymbol{\kappa}_{m}$ in the calculation. However, in certain
circumstances further approximations are possible. For example, if
the grating period $a$ (see Fig. 1b) is small enough, then for at
least some angles of incidence there will be only one propagating
diffracted order $(m=-1)$ in addition to the specularly reflected
and transmitted fields (see Fig. \ref{fig:simple_cfg}, again with
$\varepsilon_{1}=\varepsilon_{2}$). A choice of $2N+1=3$ could be
adopted, but since the field associated with $\boldsymbol{\kappa}_{1}$
is evanescent we can neglect that field and still respect energy conservation
in a lossless structure if we keep only the fields at $\boldsymbol{\kappa}_{0}$
and $\boldsymbol{\kappa}_{-1}$, simply neglecting the fields at $\boldsymbol{\kappa}_{1}$.
If we do this, and consider the simple excitation scenario presented
above, each of the $\mathbb{T}_{g,\alpha}$ and $\mathbb{R}_{g,\alpha}$
is a $2\times2$ matrix, and the resulting equations for the specularly
reflected and transmitted fields, and the diffracted fields, can be
solved easily. We refer to this as the ``two wave vector model.''
Considering an incident field from $z=-\infty$, for $s-$polarization
we find specularly transmitted and reflected fields 
\begin{align}
\frac{\mathbb{E}_{out,s}^{+}(\boldsymbol{\kappa}_{0})}{\mathbb{E}_{in,s}^{+}(\boldsymbol{\kappa}_{0})} & =U_{s}^{-1}\left(1-\frac{i\tilde{\omega}^{2}D}{2w_{1}(\boldsymbol{\kappa}_{-1})}\bar{\chi}_{00}^{\parallel}\right),\label{eq:s_spec}\\
\frac{\mathbb{E}_{out,s}^{-}(\boldsymbol{\kappa}_{0})}{\mathbb{E}_{in,s}^{+}(\boldsymbol{\kappa}_{0})} & =\frac{\mathbb{E}_{out,s}^{+}(\boldsymbol{\kappa}_{0})}{\mathbb{E}_{in,s}^{+}(\boldsymbol{\kappa}_{0})}-1,\nonumber 
\end{align}
and upward and downward diffracted fields that are equal in amplitude,
\begin{align}
\frac{\mathbb{E}_{out,s}^{\pm}(\boldsymbol{\kappa}_{-1})}{\mathbb{E}_{in,s}^{+}(\boldsymbol{\kappa}_{0})}= & U_{s}^{-1}\left(\frac{-i\tilde{\omega}^{2}D}{2w_{1}(\boldsymbol{\kappa}_{-1})}\bar{\chi}_{(-1)0}^{\parallel}\right),\label{eq:s_diff}
\end{align}
where 
\begin{align*}
U_{s} & =\left(1-\frac{i\tilde{\omega}^{2}D}{2w_{1}(\boldsymbol{\kappa}_{0})}\bar{\chi}_{00}^{\parallel}\right)\left(1-\frac{i\tilde{\omega}^{2}D}{2w_{1}(\boldsymbol{\kappa}_{-1})}\bar{\chi}_{00}^{\parallel}\right)\\
 & +\frac{\tilde{\omega}^{4}D^{2}}{4w_{1}(\boldsymbol{\kappa}_{0})w_{1}(\boldsymbol{\kappa}_{-1})}\bar{\chi}_{0(-1)}^{\parallel}\bar{\chi}_{(-1)0}^{\parallel},
\end{align*}
while for $p-$polarization we find specularly transmitted and reflected
fields 
\begin{align}
 & \frac{\mathbb{E}_{out,p}^{+}(\boldsymbol{\kappa}_{0})}{\mathbb{E}_{in,p}^{+}(\boldsymbol{\kappa}_{0})}\label{eq:p_spec}\\
 & =U_{z}^{-1}\left(1-\frac{i\kappa_{-1}^{2}D}{2w_{1}(\boldsymbol{\kappa}_{-1})\varepsilon_{1}}\bar{\chi}_{00}^{\perp}\right)+U_{\kappa}^{-1}\left(1-\frac{iw_{1}(\boldsymbol{\kappa}_{-1})D}{2\varepsilon_{1}}\bar{\chi}_{00}^{\parallel}\right)-1,\nonumber \\
 & \frac{\mathbb{E}_{out,p}^{-}(\boldsymbol{\kappa}_{0})}{\mathbb{E}_{in,p}^{+}(\boldsymbol{\kappa}_{0})}\nonumber \\
 & =U_{z}^{-1}\left(1-\frac{i\kappa_{-1}^{2}D}{2w_{1}(\boldsymbol{\kappa}_{-1})\varepsilon_{1}}\bar{\chi}_{00}^{\perp}\right)-U_{\kappa}^{-1}\left(1-\frac{iw_{1}(\boldsymbol{\kappa}_{-1})D}{2\varepsilon_{1}}\bar{\chi}_{00}^{\parallel}\right),\nonumber 
\end{align}
and upward and downward diffracted fields
\begin{align}
 & \frac{\mathbb{E}_{out,p}^{\pm}(\boldsymbol{\kappa}_{-1})}{\mathbb{E}_{in,p}^{+}(\boldsymbol{\kappa}_{0})}\label{eq:p_diff}\\
 & =U_{z}^{-1}\left(\frac{i\kappa_{0}\kappa_{-1}D}{2w_{1}(\boldsymbol{\kappa}_{-1})\varepsilon_{1}}\bar{\chi}_{(-1)0}^{\perp}\right)\pm U_{\kappa}^{-1}\left(\frac{iw_{1}(\boldsymbol{\kappa}_{0})D}{2\varepsilon_{1}}\bar{\chi}_{(-1)0}^{\parallel}\right),\nonumber 
\end{align}
 where 
\begin{align*}
U_{\kappa} & =\left(1-\frac{iw_{1}(\boldsymbol{\kappa}_{0})D}{2\varepsilon_{1}}\bar{\chi}_{00}^{\parallel}\right)\left(1-\frac{iw_{1}(\boldsymbol{\kappa}_{-1})D}{2\varepsilon_{1}}\bar{\chi}_{00}^{\parallel}\right)\\
 & +\frac{w_{1}(\kappa_{0})w_{1}(\boldsymbol{\kappa}_{-1})D^{2}}{4\varepsilon_{1}^{2}}\bar{\chi}_{0(-1)}^{\parallel}\bar{\chi}_{(-1)0}^{\parallel},\\
U_{z} & =\left(1-\frac{i\kappa_{0}^{2}D}{2w_{1}(\boldsymbol{\kappa}_{0})\varepsilon_{1}}\bar{\chi}_{00}^{\perp}\right)\left(1-\frac{i\kappa_{-1}^{2}D}{2w_{1}(\boldsymbol{\kappa}_{-1})\varepsilon_{1}}\bar{\chi}_{00}^{\perp}\right)\\
 & +\frac{\kappa_{0}^{2}\kappa_{-1}^{2}D^{2}}{4\varepsilon_{1}^{2}w_{1}(\boldsymbol{\kappa}_{0})w_{1}(\boldsymbol{\kappa}_{-1})}\bar{\chi}_{0(-1)}^{\perp}\bar{\chi}_{(-1)0}^{\perp}.
\end{align*}
Of course, the diffracted fields only appear for $\kappa_{-1}<\tilde{\omega}n_{1}$,
and the expressions above are to be used only in that range. The more
complicated form of the results for $p-$polarized light arises because
of the two components ($\hat{\boldsymbol{\kappa}}$ and $\hat{\boldsymbol{z}})$
of the light that arise, as opposed to the simpler results for $s-$polarization
where there is only one component ($\hat{\boldsymbol{s}}).$ 

\begin{figure}
\centering

\includegraphics[width=0.98\columnwidth]{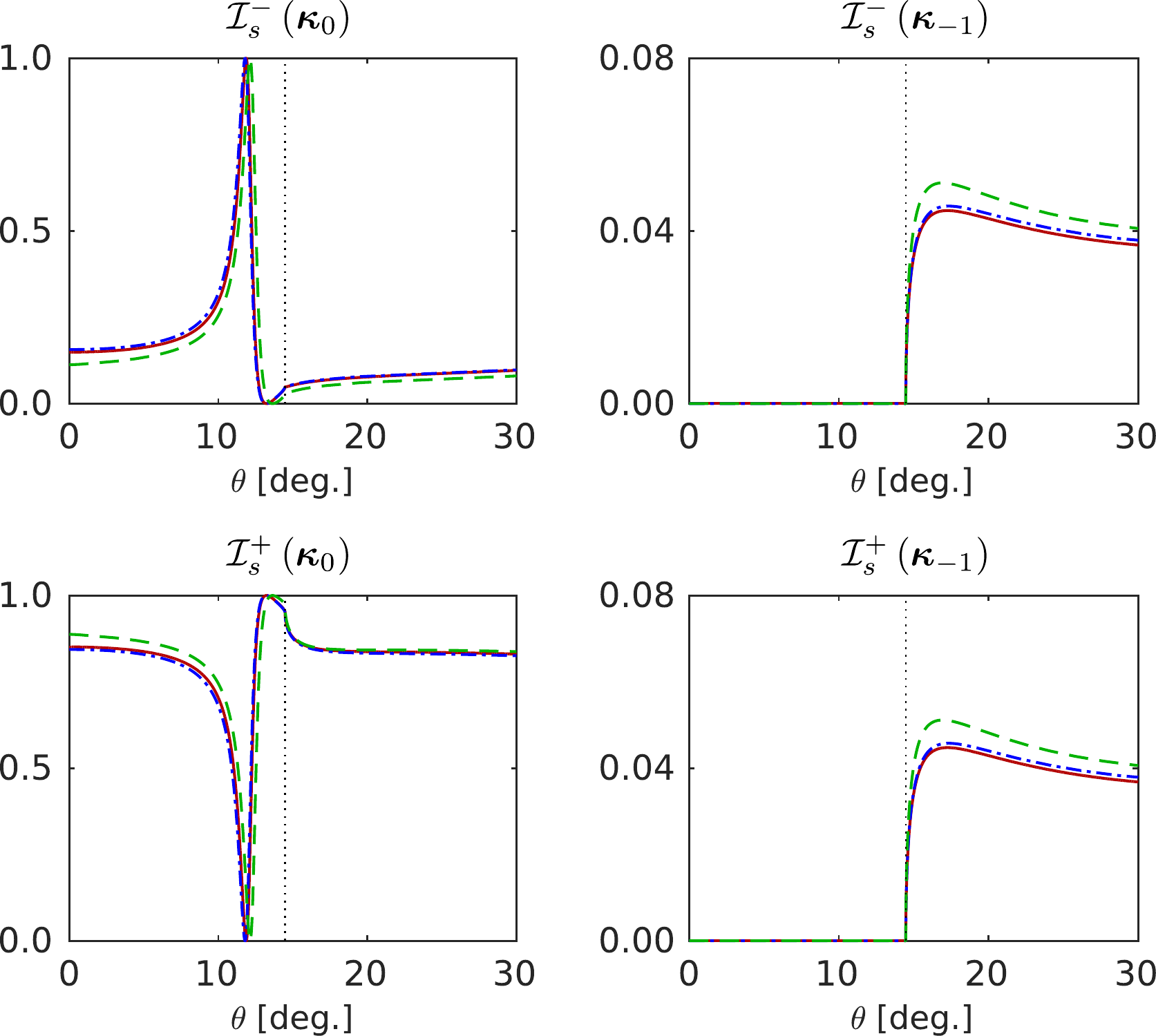}

\caption{(Color online) Comparison between the numerically exact relative irradiances
(red solid curves), the irradiances predicted by the full thin grating
model with $2N+1=7$ (blue dash-dot curves) and the irradiances predicted
from the two wave-vector model (green dashed curves). The calculation
is for a $25\,{\rm nm}$ thick grating with $a=1.24\,\mu{\rm m}$,
$d=0.62\,\mu{\rm m}$, and a refractive index of $n_{g}=3.5$ suspended
in vacuum and subject to an $s$-polarized field incident at angle
$\theta$ at a wavelength of $1.55\,\mu{\rm m}$. The vertical, dotted,
black line marks the Rayleigh anomaly.}

\label{fig:anlytic_scomp}
\end{figure}

As an example, we consider a grating with thickness $D=25\,{\rm nm}$
consisting of an isotropic medium with refractive index $3.5$ embedded
in vacuum; we take the grating period to be $a=1.25\,{\rm \mu m}$,
with a fill fraction of one-half ($d/a=0.5)$, and consider incident
$s-$polarized light at a wavelength of $1.55\,{\rm \mu m}$. In Fig.
\ref{fig:anlytic_scomp} we plot the relative irradiance of the radiated
electric fields in this system, 
\begin{align}
\mathcal{I}_{s}^{+}\left(\boldsymbol{\kappa}_{m}\right) & =\frac{\left|\mathbb{E}_{out,s}^{+}\left(\boldsymbol{\kappa}_{m}\right)\right|^{2}n_{1}\cos\theta_{1}\left(\boldsymbol{\kappa}_{m}\right)}{\left|\mathbb{E}_{in,s}^{+}\left(\boldsymbol{\kappa}_{0}\right)\right|^{2}n_{Q}\cos\theta},\label{eq:irradiance}\\
\mathcal{I}_{s}^{-}\left(\boldsymbol{\kappa}_{m}\right) & =\frac{\left|\mathbb{E}_{out,s}^{-}\left(\boldsymbol{\kappa}_{m}\right)\right|^{2}\cos\theta_{Q}\left(\boldsymbol{\kappa}_{m}\right)}{\left|\mathbb{E}_{in,s}^{+}\left(\boldsymbol{\kappa}_{0}\right)\right|^{2}\cos\theta},\nonumber 
\end{align}
for $\kappa_{m}<\tilde{\omega}n_{1}$, and $\mathcal{I}_{s}^{\pm}\left(\boldsymbol{\kappa}_{m}\right)=0$
for $\kappa_{m}>\tilde{\omega}n_{1}$ where the fields are evanescent.
Here $\theta$ is the angle of incidence, $\cos\theta=w_{Q}\left(\boldsymbol{\kappa}_{0}\right)/\tilde{\omega}n_{Q}$,
$\cos\theta_{1}\left(\boldsymbol{\kappa}_{m}\right)=w_{1}\left(\boldsymbol{\kappa}_{m}\right)/\tilde{\omega}n_{1}$,
and $\cos\theta_{Q}\left(\boldsymbol{\kappa}_{m}\right)=w_{Q}\left(\boldsymbol{\kappa}_{m}\right)/\tilde{\omega}n_{Q}$.
In green (dashed lines) we show the predictions of the specularly
and diffracted reflectance and transmission as a function of incident
angle $\theta$ (see Fig. \ref{fig:simple_cfg}) for the two wave-vector
model (\ref{eq:s_spec},\ref{eq:s_diff}); we plot in blue (dash-dot)
the predictions of the full thin grating model (\ref{eq:scat_simple})
with $2N+1=7$; and we plot in red (solid) the predictions of a full
numerical calculation using the approach of Whittaker and Culshaw
\cite{Whittaker1999,Liscidini2008}, which can be considered exact.
We see that even our simple analytic two-wave vector model (\ref{eq:s_spec},\ref{eq:s_diff})
gives a very good approximation of the diffracted and specularly reflected
and transmitted fields, and the calculation with $(2N+1)=7$ wave
vectors is essentially exact. Similar good agreement between the approximate
calculations and the numerically exact calculation is found for $p$-polarized
light.

The vertical, dotted, black lines in Fig. \ref{fig:anlytic_scomp}
identify the onset of diffraction, and thus the angle at which the
Rayleigh anomalies appear in the specularly reflected and transmitted
fields. At lower angles is the Wood anomaly: The peak in the specularly
reflected intensity, and the dip in the specularly transmitted intensity,
arise from a pole in the response functions of the structure; the
pole is associated with the ``effective waveguide'' discussed in
section III. Returning to the full response equations (\ref{eq:scatmat}),
we see that the poles of the full structure are given by 

\begin{equation}
\det\left[\bar{1}_{3}-\epsilon_{0}\bar{g}\left(\bar{\chi}_{o}+\bar{\chi}_{v}\right)\right]=0,\label{eq:FieldPole}
\end{equation}
(compare (\ref{eq:mode},\ref{eq:mode_conditions})). Poles here are
off the real $\kappa$ axis; the $\boldsymbol{K}\neq0$ components
of the grating provide coupling into and out of the uniform waveguide,
with a dispersion relation identified approximately by the expressions
(\ref{eq:mode_conditions}), giving the position of the pole in the
$\kappa$ plane an imaginary contribution, as well as a shift in the
real component of the pole. To verify this, we restrict ourselves
to excitation with $\boldsymbol{\kappa}_{0}\cdot\hat{\boldsymbol{y}}>0$
and expand our analytic expressions for the specular component of
the electric field in (\ref{eq:s_spec}) and (\ref{eq:p_spec}) about
$\kappa_{0}=\check{\kappa}$, with $\check{\kappa}$ defined by the
expression 
\begin{align*}
\check{\kappa}-\frac{2\pi}{a} & =-\kappa_{WG},
\end{align*}
where $\kappa_{WG}$ is the magnitude of the wave vector satisfying
the approximate dispersion relations of the isolated waveguide mode
given by (\ref{eq:mode_conditions}); expressions for $\kappa_{WG}$
for $s-$ and $p$-polarization are given by (\ref{eq:dispeqs}) and
(\ref{eq:dispeqp}) in Appendix \ref{sec:WGmodes}. For $\kappa_{0}$
in this region $\boldsymbol{\kappa}_{-1}\cdot\hat{\boldsymbol{y}}<0$,
and $\boldsymbol{\kappa}_{-1}=-(2\pi/a-\kappa_{0})=-\kappa_{-1}\hat{\boldsymbol{y}}$
is close to the wave vector of a waveguide mode propagating in the
$-\hat{\boldsymbol{y}}$ direction, $\boldsymbol{\kappa}_{-1}\approx\boldsymbol{\check{\kappa}}_{-1}'\equiv-(2\pi/a-\check{\kappa})\hat{\boldsymbol{y}}=-\kappa_{WG}\hat{\boldsymbol{y}}$.
Since $\kappa_{WG}>\tilde{\omega}n_{1}$, $w_{1}(\kappa_{WG})$ is
purely imaginary; we put $q=-iw_{1}(\kappa_{WG})$, use superscripts
$s$ and $p$ on $\check{\kappa},$ $\kappa_{WG}$, and $q$ to indicate
the appropriate polarization, and also use $w_{1}^{s}$ and $w_{1}^{p}$
as short-hand for $w_{1}(\check{\kappa}^{s})$ and $w_{1}(\check{\kappa}^{p}$)
respectively. Looking at the transmitted specular field, for $\boldsymbol{\kappa}_{0}$
in the neighborhood of $\check{\kappa}\hat{\boldsymbol{y}}$ we find
the expressions (\ref{eq:s_spec},\ref{eq:p_spec}) can be written
approximately as 
\begin{align}
\frac{\mathbb{E}_{out,s}^{+}(\boldsymbol{\kappa}_{0})}{\mathbb{E}_{in,s}^{+}(\boldsymbol{\kappa}_{0})} & \approx\eta_{s}\frac{\kappa_{-1}-\kappa_{WG}}{\kappa_{-1}-(\kappa_{R}^{s}+i\kappa_{I}^{s})},\label{eq:analytic_expanded}\\
\frac{\mathbb{E}_{out,p}^{+}(\boldsymbol{\kappa}_{0})}{\mathbb{E}_{in,p}^{+}(\boldsymbol{\kappa}_{0})} & \approx\eta_{p}\frac{\kappa_{-1}-\kappa_{WG}}{\kappa_{-1}-(\kappa_{R}^{p}+i\kappa_{I}^{p})}+C,\nonumber 
\end{align}
where $\kappa_{R}^{s,p}\equiv\kappa_{WG}^{s,p}+\kappa_{\delta}^{s,p}$,
with 
\begin{align*}
\kappa_{\delta}^{s} & =-\frac{\left(q^{s}\right)^{2}}{\kappa_{WG}^{s}}\left(\frac{\tilde{\omega}^{2}D}{2w_{1}^{s}}\right)^{2}\frac{\bar{\chi}_{0(-1)}^{\parallel}\bar{\chi}_{(-1)0}^{\parallel}}{1+\left(\frac{\tilde{\omega}^{2}D}{2w_{1}^{s}}\bar{\chi}_{00}^{\parallel}\right)^{2}},\\
\kappa_{I}^{s} & =\frac{\left(q^{s}\right)^{2}}{\bar{\chi}_{00}^{\parallel}\kappa_{WG}^{s}}\left(\frac{\tilde{\omega}^{2}D}{2w_{1}^{s}}\right)\frac{\bar{\chi}_{0(-1)}^{\parallel}\bar{\chi}_{(-1)0}^{\parallel}}{1+\left(\frac{\tilde{\omega}^{2}D}{2w_{1}^{s}}\chi_{00}^{\parallel}\right)^{2}},\\
\eta_{s} & =\frac{1}{1-\frac{i\tilde{\omega}^{2}D}{2w_{1}^{s}}\bar{\chi}_{00}^{\parallel}},
\end{align*}
and

\begin{align*}
\kappa_{\delta}^{p} & =-\left(\frac{\left(\check{\kappa}^{p}\right)^{2}D}{2\varepsilon_{1}w_{1}^{p}}\right)^{2}\frac{\kappa_{WG}^{p}}{\left[\left(\kappa_{WG}^{p}\right)^{2}/\left(q^{p}\right)^{2}-2\right]}\frac{\bar{\chi}_{0(-1)}^{\perp}\bar{\chi}_{(-1)0}^{\perp}}{\left[1+\left(\frac{\left(\check{\kappa}^{p}\right)^{2}D}{2\varepsilon_{1}w_{1}^{p}}\bar{\chi}_{00}^{\perp}\right)^{2}\right]},\\
\kappa_{I}^{p} & =\frac{1}{\bar{\chi}_{00}^{\perp}}\left(\frac{\left(\check{\kappa}^{p}\right)^{2}D}{2\varepsilon_{1}w_{1}^{p}}\right)\frac{\kappa_{WG}^{p}}{\left[\left(\kappa_{WG}^{p}\right)^{2}/\left(q^{p}\right)^{2}-2\right]}\frac{\bar{\chi}_{0(-1)}^{\perp}\bar{\chi}_{(-1)0}^{\perp}}{\left[1+\left(\frac{\left(\check{\kappa}^{p}\right)^{2}D}{2\varepsilon_{1}w_{1}^{p}}\bar{\chi}_{00}^{\perp}\right)^{2}\right]},\\
\eta_{p} & =\frac{1}{1-\frac{i\left(\check{\kappa}^{p}\right)^{2}D}{2\varepsilon_{1}w_{1}^{p}}\bar{\chi}_{00}^{\perp}},
\end{align*}
and where

\[
C=\frac{1}{U_{\kappa}}\left(1+\frac{q^{p}D}{2\varepsilon_{1}}\bar{\chi}_{00}^{\parallel}\right)-1
\]
is negligible for sufficiently thin gratings. We do not plot (\ref{eq:analytic_expanded}),
but note that in the region of the dip of the specular transmission
for both $s$- and $p$-polarized light the pole expansion gives an
extremely good fit to the more exact expressions (\ref{eq:s_spec},\ref{eq:p_spec})
in the two-wave-vector model, as well of course to the results (\ref{eq:TRs},\ref{eq:TRp})
of the $(2N+1)$-wave-vector model and to the exact numerical results
with which the two-wave-vector model agrees well. The inclusion of
the imaginary parts $\kappa_{I}^{s,p}$ of the pole positions are
obviously essential in achieving this, but the inclusion of the shifts
$\kappa_{\delta}^{s,p}$ in the real part of the pole positions are
as well and should not be neglected; both are second order in the
grating coupling amplitudes.

\subsection{Including a substrate\label{subsec:substrate}}

Returning to our general scattering treatment (\ref{eq:full_scattering})
of an isolated grating, we can move to a transfer matrix treatment
by solving for the upward and downward propagating (or evanescent)
field amplitudes above the grating ($\mathbb{\bar{E}}_{out}^{+}$
and $\mathbb{\bar{E}}_{in}^{-}$) in terms of the upward and downward
propagating (or evanescent) field amplitudes below the grating ($\mathbb{\bar{E}}_{in}^{+}$
and $\mathbb{\bar{E}}_{out}^{-}$),

\begin{equation}
\left[\begin{array}{c}
\bar{\mathbb{E}}_{out}^{+}\\
\mathbb{\bar{\mathbb{E}}}_{in}^{-}
\end{array}\right]=\bar{\mathbb{M}}_{g}\left[\begin{array}{c}
\bar{\mathbb{E}}_{in}^{+}\\
\bar{\mathbb{E}}_{out}^{-}
\end{array}\right],
\end{equation}
where 
\begin{equation}
\bar{\mathbb{M}}_{g}=\left[\begin{array}{cc}
\bar{\mathbb{T}}_{g}-\bar{\mathbb{R}}_{g}\left(\bar{\mathbb{T}}_{g}\right)^{-1}\bar{\mathbb{R}}_{g} & \bar{\mathbb{R}}_{g}\left(\bar{\mathbb{T}}_{g}\right)^{-1}\\
-\left(\bar{\mathbb{T}}_{g}\right)^{-1}\bar{\mathbb{R}}_{g} & \left(\bar{\mathbb{T}}_{g}\right)^{-1}
\end{array}\right]\label{eq:Mgrat}
\end{equation}
has $4(2N+1)\times4(2N+1)$ elements, as does the scattering matrix
(\ref{eq:S}). Returning to our general structure of Fig.  \subref*{fig:grating_multi}, we
can now combine the transfer matrix (\ref{eq:Mgrat}) of the grating
region with the transfer matrix of the multilayer below to form a
transfer matrix for the whole structure, in terms of which the optical
properties of the structure can be calculated. To do this, consider
first light characterized by a single \textbf{$\boldsymbol{\kappa}$}
in the presence of the multilayer structure of Fig.  \subref*{fig:grating_multi}, but $without$
the presence of the grating. The transfer matrix of the multilayer
structure relating upward- and downward propagating (or evanescent)
amplitudes of light just above the multilayer in the medium with relative
dielectric constant $\varepsilon_{1}$ (at $z=\left(-D/2\right)^{+})$
to upward- and downward propagating (or evanescent) amplitudes of
light at the largest $z$ to which the substrate, with relative dielectric
constant $\varepsilon_{Q}$, extends, takes the form
\begin{equation}
\mathbb{M}_{1Q}=\begin{bmatrix}\mathbb{T}_{Q1}-\mathbb{R}_{1Q}\left(\mathbb{T}_{1Q}\right)^{-1}\mathbb{R}_{Q1} & \mathbb{R}_{1Q}\left(\mathbb{T}_{1Q}\right)^{-1}\\
-\left(\mathbb{T}_{1Q}\right)^{-1}\mathbb{R}_{Q1} & \left(\mathbb{T}_{1Q}\right)^{-1}
\end{bmatrix}.\label{eq:Msub_nodiff}
\end{equation}
This is a $4\times4$ matrix, but as long as the layered materials
are isotropic or uniaxial it will be composed of $2\times2$ block
matrices $\mathbb{T}_{ij}=diag\left(\mathbb{T}_{ij}^{s},\mathbb{T}_{ij}^{p}\right)$
where $\mathbb{T}_{ij}^{s,p}$ is the Fresnel coefficient for the
transmitted $s-$ or $p-$polarized fields from $\varepsilon_{i}$
to $\varepsilon_{j}$ and $\mathbb{R}_{ij}$ is similarly defined
for their reflected counterparts \cite{Sipe1987}. We can immediately
extend this to a transfer matrix $\mathbb{\bar{M}}_{1Q}$ of the layered
structure involving all our $(2N+1)$ $\boldsymbol{\kappa}_{m}$ of
interest by writing 

\begin{align}
\bar{\mathbb{M}}_{1Q} & =\begin{bmatrix}\bar{\mathbb{T}}_{Q1}-\bar{\mathbb{R}}_{1Q}\left(\bar{\mathbb{T}}_{1Q}\right)^{-1}\bar{\mathbb{R}}_{Q1} & \bar{\mathbb{R}}_{1Q}\left(\bar{\mathbb{T}}_{1Q}\right)^{-1}\\
-\left(\bar{\mathbb{T}}_{1Q}\right)^{-1}\bar{\mathbb{R}}_{Q1} & \left(\bar{\mathbb{T}}_{1Q}\right)^{-1}
\end{bmatrix},\label{eq:Msub}
\end{align}
a $4(2N+1)\times4(2N+1)$ matrix where $\left[\bar{\mathbb{T}}_{1Q}\right]_{mm}=diag\left(\mathbb{T}_{1Q}^{s}\left(\kappa_{m}\right),\mathbb{T}_{1Q}^{p}\left(\kappa_{m}\right)\right)$
and other terms are similarly defined. 

We can now construct a transfer matrix for the full structure shown
in Fig.  \subref*{fig:grating_multi} by imagining an infinitesimal layer of material with
relative dielectric constant $\varepsilon_{1}$ inserted between the
bottom of the grating structure and the top of the highest layer in
the multilayer structure below. Then the transfer matrix relating
the upward and downward propagating (or evanescent) field amplitudes
just above the grating to the upward and downward propagating (or
evanescent) field amplitudes at the largest $z$ in the substrate
is given by

\[
\bar{\mathbb{M}}_{1Q}^{\prime}=\bar{\mathbb{M}}_{g}\bar{\mathbb{M}}_{1}\left(D/2\right)\bar{\mathbb{M}}_{1Q},
\]
where 
\[
\bar{\mathbb{M}}_{1}\left(D/2\right)=\begin{bmatrix}\bar{\mathbb{L}}^{+} & \bar{0}\\
\bar{0} & \bar{\mathbb{L}}^{-}
\end{bmatrix}
\]
is composed of $2\left(2N+1\right)\times2\left(2N+1\right)$ block
matrices with $\bar{\mathbb{L}}^{\pm}\left(\boldsymbol{\kappa}_{m}\right)=e^{\pm iw_{1}\left(\boldsymbol{\kappa}_{m}\right)D/2}\begin{bmatrix}1 & 0\\
0 & 1
\end{bmatrix}$ and propagates the fields from the center of the grating at $z=0$
to the position of the substrate at $z=-D/2$. Through simple algebra
we can write the elements of $\bar{\mathbb{M}}_{1Q}^{\prime}$ as

\begin{figure*}[ht]
\centering
\includegraphics[width=0.8\textwidth]{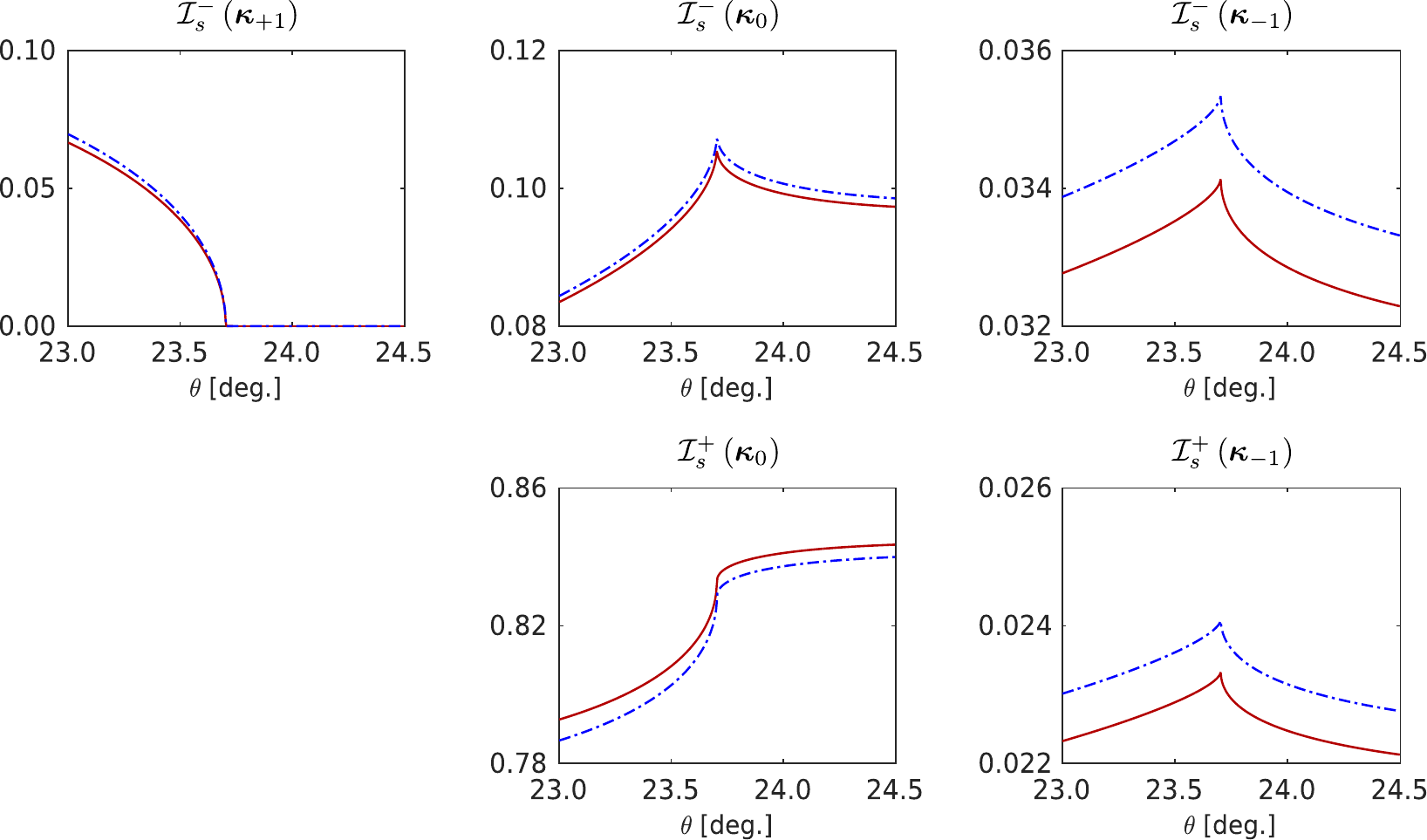}
\caption{(Color online) Comparison between relative irradiance calculations that are numerically exact (red solid curves) and those predicted by the full thin grating model with $2N+1=7$ (blue dash-dot curves) around a Rayleigh anomaly. The calculation is for a $25\,{\rm nm}$ thick grating with $a=1.8\,\mu{\rm m}$, $d=0.72\,{\rm \mu m}$, and a refractive index of $n_{g}=3.5$. The system has a substrate with index $n_{Q}=1.44$, a vacuum cladding, and is subject to an $s$-polarized incident field from the substrate at angle $\theta$ and with a vacuum wavelength of $1.55\,\mu{\rm m}$.}

\label{fig:nc1_Rayleigh}

\end{figure*}

\begin{align}
\bar{\mathbb{M}}_{1Q}^{\prime} & =\begin{bmatrix}\bar{\mathbb{T}}_{Q1}^{\prime}-\bar{\mathbb{R}}_{1Q}^{\prime}\left(\bar{\mathbb{T}}_{1Q}^{\prime}\right)^{-1}\bar{\mathbb{R}}_{Q1}^{\prime} & \bar{\mathbb{R}}_{1Q}^{\prime}\left(\bar{\mathbb{T}}_{1Q}^{\prime}\right)^{-1}\\
-\left(\bar{\mathbb{T}}_{1Q}^{\prime}\right)^{-1}\bar{\mathbb{R}}_{Q1}^{\prime} & \left(\bar{\mathbb{T}}_{1Q}^{\prime}\right)^{-1}
\end{bmatrix},\label{eq:Mpr}
\end{align}
where
\begin{eqnarray}
\bar{\mathbb{T}}_{1Q}^{\prime} & = & \bar{\mathbb{T}}_{1Q}\bar{\mathbb{L}}^{+}\left[\bar{1}_{2}-\bar{\mathbb{R}}_{g}\bar{\mathbb{R}}_{1Q}\left(\bar{\mathbb{L}}^{+}\right)^{2}\right]^{-1}\bar{\mathbb{T}}_{g},\nonumber \\
\bar{\mathbb{T}}_{Q1}^{\prime} & = & \bar{\mathbb{T}}_{g}\bar{\mathbb{R}}_{1Q}\left(\bar{\mathbb{L}}^{+}\right)^{2}\left[\bar{1}_{2}-\bar{\mathbb{R}}_{g}\bar{\mathbb{R}}_{1Q}\left(\bar{\mathbb{L}}^{+}\right)^{2}\right]^{-1}\left(\bar{\mathbb{R}}_{1Q}\right)^{-1}\bar{\mathbb{T}}_{Q1}\bar{\mathbb{L}}^{-},\nonumber \\
\bar{\mathbb{R}}_{1Q}^{\prime} & = & \bar{\mathbb{R}}_{g}+\bar{\mathbb{T}}_{g}\bar{\mathbb{R}}_{1Q}\left(\bar{\mathbb{L}}^{+}\right)^{2}\left[\bar{1}_{2}-\bar{\mathbb{R}}_{g}\bar{\mathbb{R}}_{1Q}\left(\bar{\mathbb{L}}^{+}\right)^{2}\right]^{-1}\bar{\mathbb{T}}_{g},\label{eq:Fresnelprime}\\
\bar{\mathbb{R}}_{Q1}^{\prime} & = & \bar{\mathbb{R}}_{Q1}+\bar{\mathbb{T}}_{1Q}\bar{\mathbb{L}}^{+}\left[\bar{1}_{2}-\bar{\mathbb{R}}_{g}\bar{\mathbb{R}}_{1Q}\left(\bar{\mathbb{L}}^{+}\right)^{2}\right]^{-1}\bar{\mathbb{R}}_{g}\bar{\mathbb{T}}_{Q1}\bar{\mathbb{L}}^{+}\nonumber 
\end{eqnarray}
are easily identified as the transmission and reflection matrices
of the entire structure (compare (\ref{eq:Mgrat},\ref{eq:Msub})).

The analytic structure of the new Fresnel matrices (\ref{eq:Fresnelprime})
is inherited from that of the isolated grating (\ref{eq:scatmat})
and from the $\mathbb{\bar{R}}_{ij}$ and $\mathbb{\bar{T}}_{ij}$
of the multilayer below it. Besides the poles of $\mathbb{\bar{T}}_{g}$
and $\mathbb{\bar{R}}_{g}$ signaling the waveguide modes in the isolated
grating structure, we can in general expect poles in $\mathbb{\bar{R}}_{ij}$
and $\mathbb{\bar{T}}_{ij}$ signaling the presence of waveguide modes
in the multilayer. The positions of the poles in the new Fresnel matrices
(\ref{eq:Fresnelprime}) will exhibit the interaction between these
excitations, and we will turn to a general analysis of the new excitations
in a later publication. Here we focus on a first application our thin
grating model in the presence of a substrate, and on some of the qualitative
features that arise from the interaction. Thus we consider the simplest
multilayer structure possible, taking the substrate with relative
dielectric constant $\varepsilon_{Q}$ to extend up to $z=-D/2$.
The grating, suspended in a medium of dielectric constant $\varepsilon_{1}$
which also serves as a cladding, then resides on a semi-infinite substrate
of dielectric constant $\varepsilon_{Q}$, which we now relabel $\varepsilon_{2}$
(see Fig. \ref{fig:simple_cfg}). Taking $\varepsilon_{2}$ to be
real and positive there are no modes associated with the substrate,
and so the only effect of the substrate will be to modify the modes
identified by the poles of the isolated grating structure (see Fig.
 \subref*{fig:grating_def}).

\begin{figure*}[ht!]
\centering
\includegraphics[width=0.8\textwidth]{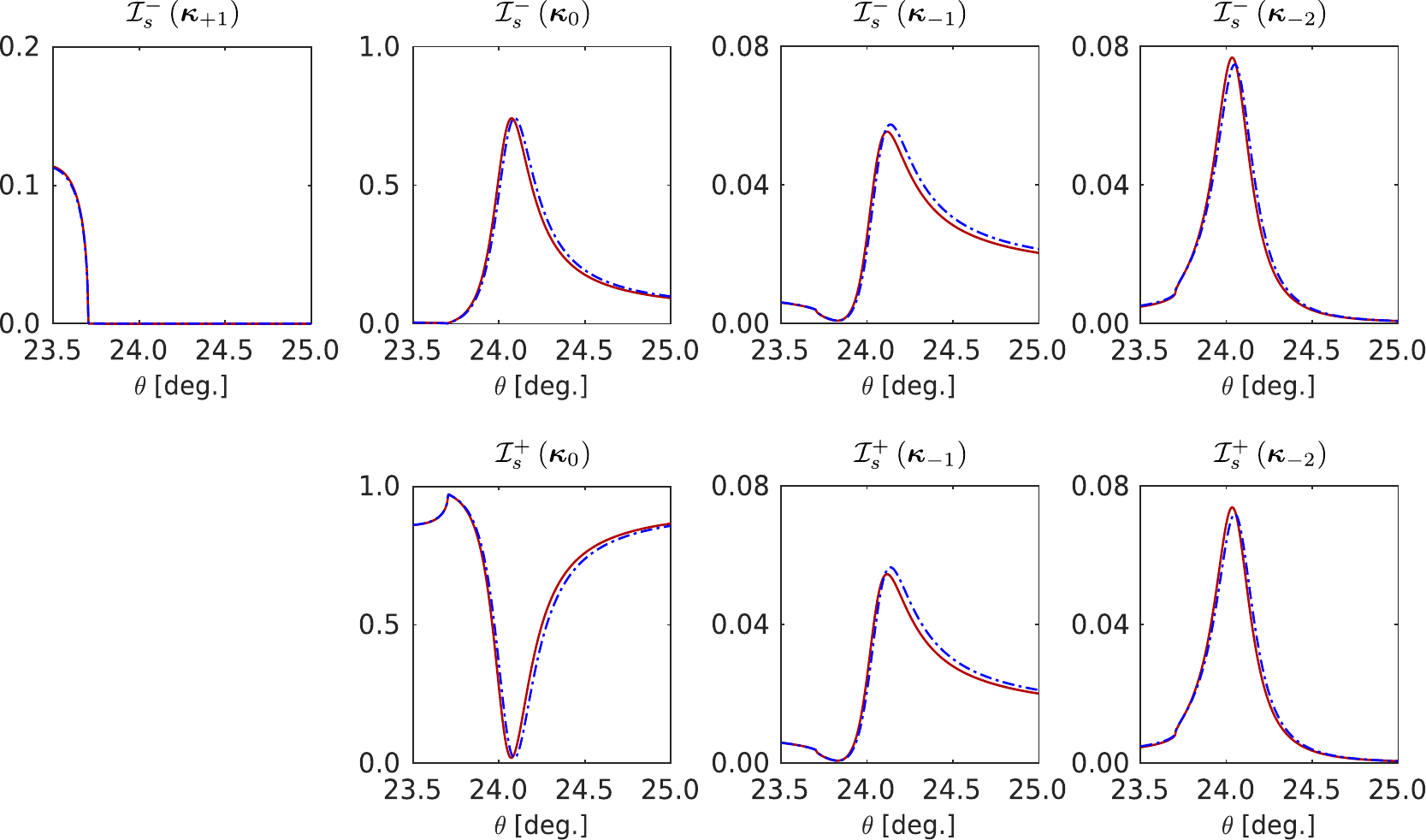}
\caption{(Color online) Comparison of relative irradiance calculations that are numerically exact (red solid curves) and those predicted by the full thin grating model (blue dash dot curves). The calculation is for a $25\,{\rm nm}$ thick grating with $a=1.8\,\mu{\rm m}$, $d=0.72\,{\rm \mu m}$, and a refractive index of $n_{g}=3.5$. The system has a substrate with index $n_{Q}=1.44$, a cladding with index $n_{1}=1.42$, and is subject to an $s-$polarized field, incident from the substrate at angle $\theta$ with a vacuum wavelength $1.55\,\mu{\rm m}$.}

\label{fig:benchplots}

\end{figure*}

Insight into the nature of this modification can be gleaned from recalling
the simplest picture of the grating region as an effective anisotropic
slab (see Fig. \subref*{fig:eff_slab}). In a symmetric environment both $s-$ and $p-$polarized
waveguide modes exist, but for an environment with $\varepsilon_{2}\neq\varepsilon_{1}$
(see Fig. \ref{fig:simple_cfg}) the modes will not survive if the
asymmetry is large enough. In such a situation we expect the Wood
anomalies will vanish, although of course the Rayleigh anomalies will
remain. We demonstrate how our thin grating model describes this situation
by considering a grating with $D=25\,{\rm nm}$, with a dielectric
constant $\varepsilon_{g}=(3.5)^{2}$ appropriate for silicon, a period
of $a=1.8\,{\rm \mu m}$, and a fill fraction $d/a=0.4$ in vacuum
($\varepsilon_{1}=1)$, located above a substrate of fused silica
($\varepsilon_{2}=(1.44)^{2})$ and subject to $s-$polarized light
from below at a vacuum wavelength of $\lambda=1.55\,{\rm \mu m}.$
For the resulting $\varepsilon_{layer}$ we would require an $\varepsilon_{1}>(1.38)^{2}$
for a waveguide mode to be contained within the guiding layer, so
the asymmetry here is too great to allow for Wood anomalies, and only
Rayleigh anomalies should survive. Assuming $\varepsilon_{2}>\varepsilon_{1}$,
this can be confirmed by solving (\ref{eq:exact_wgcon}) for $\varepsilon_{1}$
with $\kappa=\tilde{\omega}\sqrt{\varepsilon_{2}}$. In agreement
with this simple argument, the reflected, transmitted, and diffracted
light intensities exhibits only cusp-like Rayleigh anomalies, as seen
in Fig. \ref{fig:nc1_Rayleigh}. In blue (dashed-dot) we plot a calculation
with $(2N+1)=7$ wave vectors using (\ref{eq:full_scattering}), while
in red (solid) we plot the exact result found numerically from the
approach of Whittaker and Culshaw \cite{Whittaker1999,Liscidini2008}.
There is excellent qualitative and good quantitative agreement between
the results of the thin grating model and the exact result, especially
considering that a parameter $2\pi\sqrt{\varepsilon_{g}}D/\lambda$
, which should obviously be small for our thin grating approximations
(\ref{eq:uniformity}) to be valid, is here about $0.35$. There is
only a significant relative correction in the diffracted intensities
at $\boldsymbol{\kappa}_{-1}$, where the diffracted intensities themselves
are very small.

\begin{figure*}[ht]
\centering
\includegraphics[width=0.8\textwidth]{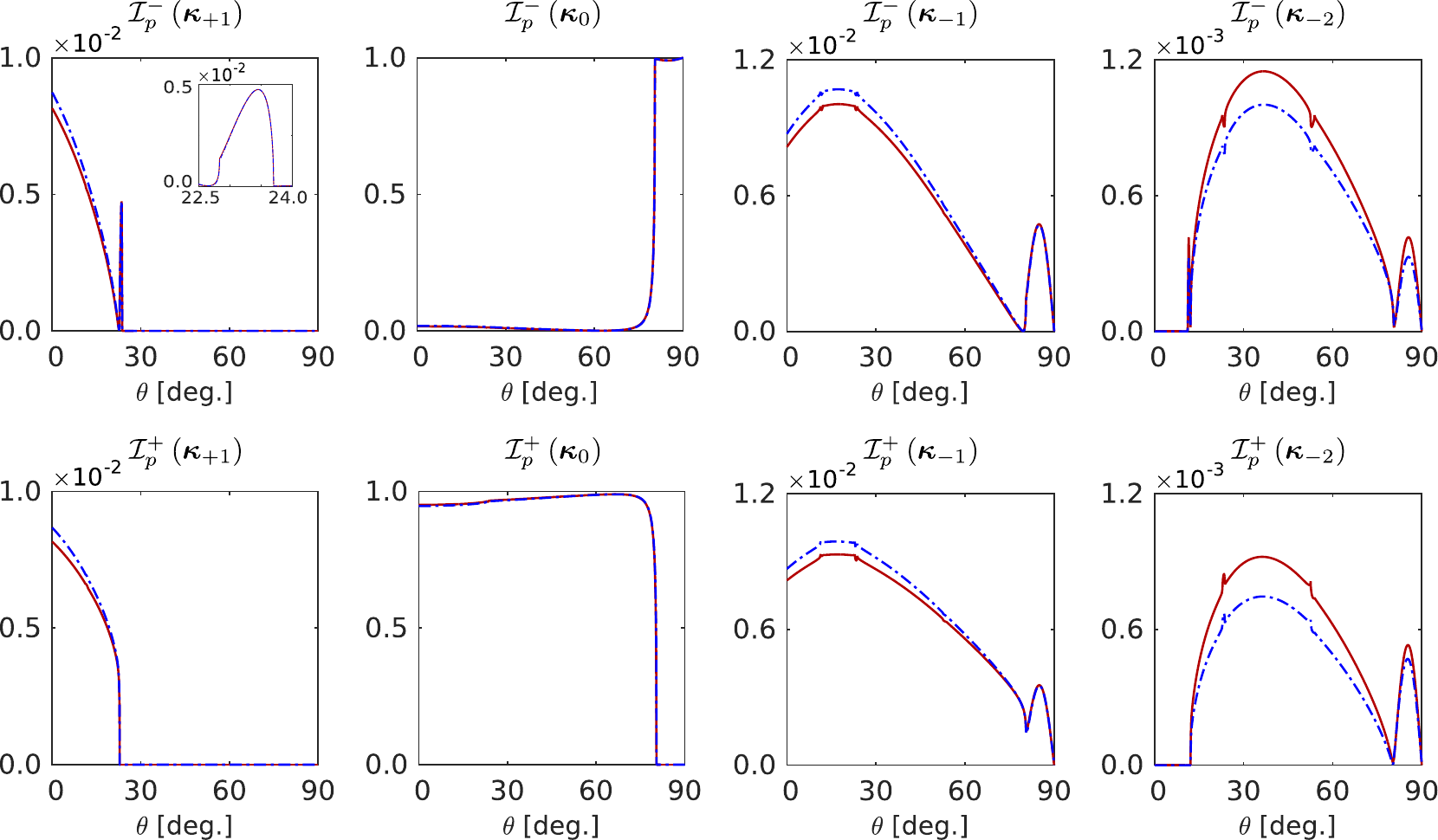}
\caption{(Color online) Comparison of relative irradiance calculations that are numerically exact (red solid curves) and those predicted by the full thin grating model (blue dash dot curves). The calculation is for a $25\,{\rm nm}$ thick grating with $a=1.8\,\mu{\rm m}$, $d=0.72\,{\rm \mu m}$, and a refractive index of $n_{g}=3.5$. The system has a substrate with index $n_{Q}=1.44$, a cladding with index $n_{1}=1.42$, and is subject to a $p-$polarized field, incident from the substrate at angle $\theta$ with a vacuum wavelength $1.55\,\mu{\rm m}$. The insert in the graph of $\mathcal{I}_{p}^{-}\left(\boldsymbol{\kappa}_{+1}\right)$ shows the detail around $\theta=23.5^{o}$.}

\label{fig:benchplots-p}

\end{figure*}

In order for this grating to exhibit a Wood anomaly the mismatch between
$\varepsilon_{2}$ and $\varepsilon_{1}$ must be decreased. To move
into this regime, we raise $\varepsilon_{1}$ to $(1.42)^{2}$, and
keep all other parameters the same; for this value the simple argument
used above predicts a waveguide mode for $s-$polarized light, but
not for $p-$polarized light. In accord with this, the calculated
reflected, transmitted, and diffracted light irradiances shown in
Figs. \ref{fig:benchplots} and \ref{fig:benchplots-p} exhibit Wood
and Rayleigh anomalies for $s-$polarized light, but only Rayleigh
anomalies for $p-$polarized light. Again in blue (dashed-dot) we
plot a calculation using (\ref{eq:full_scattering}) with $2N+1=7$,
while in red (solid) we give the results from a full numerical calculation
using the approach of Whittaker and Culshaw \cite{Whittaker1999,Liscidini2008}.
For $s-$polarized light we focus on the region around the Wood anomaly;
note that with the field incident from the substrate, which has a
higher index than the cladding, the forward diffracted fields become
evanescent before the backward diffracted fields. For $p-$polarized
light we plot the response for all incident angles; the absence of
a Wood anomaly leaves somewhat unremarkable results for specularly
reflected and transmitted light, but yields several noteworthy features
in the diffracted components, which can propagate up to $m=-3$. Rayleigh
anomalies when the $m=+1$, $-2$, and $-3$ diffracted orders transition
between evanescence and propagation lead to Rayleigh anomalies that
appear as non-analyticities in the $m=-1$ and $m=-2$ beams. Additionally,
for angles of incidence beyond that which would yield total internal
reflection were the grating absent, the specular reflectance does
not remain at unity. A small dip in the specular reflectance follows
the Rayleigh anomaly associated with this transition, which is compensated
by an increase in the irradiance of the remaining diffracted components.
They display peaks over this range, which finally drop to zero as
the incidence approaches grazing. We note excellent agreement between
the thin grating results (\ref{eq:full_scattering}) and the exact
calculation throughout the plots in Figs. \ref{fig:benchplots} and
\ref{fig:benchplots-p}, with the largest relative disagreements appearing
only when the intensities involved are very small.

Although not shown, we note that if the asymmetry between cladding
and substrate is decreased further so that a $p$-polarized Wood anomaly
appears, we observe a small shift between its location as predicted
by (\ref{eq:full_scattering}) and the full numerical results, which
does not occur for the $s-$polarized Wood anomaly shown in Fig. \ref{fig:benchplots}.
For both $s$- and $p-$polarized Wood anomalies, the disagreements
with the full numerical calculations increase as the dimensionless
optical thickness parameters $\tilde{D}_{s}$ and $\tilde{D}_{p}$,
given by (\ref{eq:Ds}) and (\ref{eq:Dp}) in Appendix \ref{sec:WGmodes},
approach unity. In that Appendix we show that this signals the breakdown
of our approximate treatment of the waveguide modes in the effective
anisotropic slab.

\section{Conclusions\label{sec:Conclusions}}

In this work we have presented a treatment for the optical response
of thin gratings. Although approximate, it nonetheless respects energy
conservation exactly, even if there are large exchanges of energy
between specular and diffracted fields, and between specularly transmitted
and reflected fields. These large exchanges are associated with Rayleigh
and Wood anomalies. Our Green function approach makes it easy to see
how the anomalies arise from the structure of the equations that describe
the specular and diffracted fields, with square root singularities
associated with Rayleigh anomalies and poles with Wood anomalies;
the poles are linked to effective waveguide modes of the grating region
that are easily identified in the thin grating limit. This helps in
understanding the optical response even if a set of coupled wave vector
equations must be solved for the specular and diffracted fields. Yet,
where only a few wave vectors are important, analytic expressions
can be given directly for the specular and diffracted fields. Comparison
with full numerical solutions of a 1D grating response confirms that
our approximate solutions is in excellent agreement with the exact
response, even near the anomalies.

We expct that the development of approximate yet accurate treatments
of thin gratings, such as the one presented here, will play an important
role in enabling their use as probes of optical systems. The calculations
can be made more easily than full numerical treatments, and the physics
can be identified in the reasonably simple sets of equations that
are used in calculations.
\begin{acknowledgments}
The authors thank the Natural Sciences and Engineering Research Council
of Canada (NSERC) for partial funding of this work including the award
of an Alexander Graham Bell Canada graduate scholarship to D. A. Travo.
Additionally, we thank Matteo Galli, Sharon Weiss, and Daniele Aurelio
for useful conversations throughout this work.
\end{acknowledgments}

\appendix

\section{Fourier components for a rectangular grating}

\label{sec:effindex}

In this section we evaluate $\boldsymbol{\chi}_{v[m]}$ for a one
dimensional rectangular grating, composed of isotropic dielectric
materials, such as the one presented in Fig.  \subref*{fig:grating_def}. Recalling (\ref{eq:chiv_def}),
we have
\begin{align}
\boldsymbol{\chi}_{v[m]} & =\frac{1}{a}\int_{-a/2}^{a/2}e^{-imK\zeta}\left(\boldsymbol{\chi}_{mod}\left(\zeta\right)-\left\langle \boldsymbol{\chi}_{mod}\right\rangle \right)d\zeta,\label{eq:Chi_vm}
\end{align}
where from (\ref{eq:Chi_eff}) and our definition of $\boldsymbol{\varepsilon}\left(\zeta\right)\equiv\boldsymbol{\varepsilon}_{1}+\boldsymbol{\chi}_{add}\left(\zeta\right)$,
we can write

\begin{align*}
\chi_{mod}^{\parallel}\left(\zeta\right) & =\varepsilon^{\parallel}\left(\zeta\right)-\varepsilon_{1},\\
\chi_{mod}^{\perp}\left(\zeta\right) & =\varepsilon_{1}\left(1-\frac{\varepsilon_{1}}{\varepsilon^{\perp}\left(\zeta\right)}\right).
\end{align*}

Note that for the system considered, we have
\[
\varepsilon^{\perp}\left(\zeta\right)=\varepsilon^{\parallel}\left(\zeta\right)=\begin{cases}
\varepsilon_{g} & \left|\zeta\right|\le d/2\\
\varepsilon_{1} & \left|\zeta\right|>d/2
\end{cases}
\]
within a single period of our grating. Additionally, we note that
$\left\langle \boldsymbol{\chi}_{mod}\right\rangle $, found from
(\ref{eq:Chi_layer}), yields $\left\langle \chi_{mod}^{\parallel}\right\rangle =\varepsilon_{layer}^{\parallel}-\varepsilon_{1}$
and $\left\langle \chi_{mod}^{\perp}\right\rangle =\varepsilon_{1}\left(1-\varepsilon_{1}/\varepsilon_{layer}^{\perp}\right)$
where $\varepsilon_{layer}^{\parallel}$ and $\varepsilon_{layer}^{\perp}$
are found from (\ref{eq:epsilon_layer}). At this point we can evaluate
(\ref{eq:Chi_vm}) to find
\begin{align*}
\chi_{v[m]}^{\parallel} & =\left(\varepsilon_{layer}^{\parallel}-\varepsilon_{1}\right)\text{sinc}\left(\frac{mKd}{2}\right),\\
\chi_{v[m]}^{\perp} & =\varepsilon_{1}\left(1-\frac{\varepsilon_{1}}{\varepsilon_{layer}^{\perp}}\right)\text{sinc}\left(\frac{mKd}{2}\right).
\end{align*}

\section{Waveguide Modes}

\label{sec:WGmodes}

\begin{figure}
\centering

\subfloat[$s$-polarization]{\includegraphics[width=0.98\columnwidth]{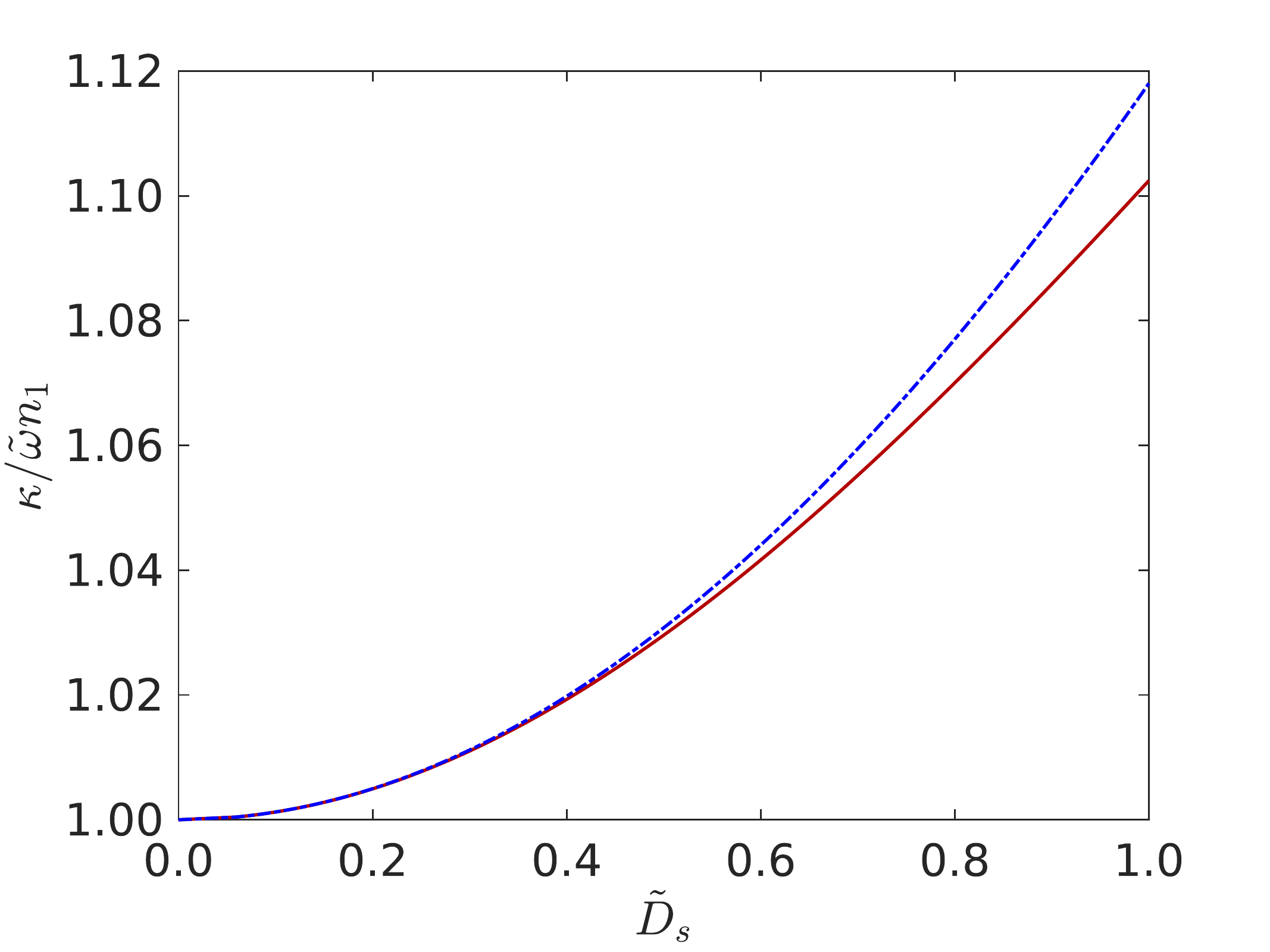}
}

\subfloat[$p$-polarization]{\includegraphics[width=0.98\columnwidth]{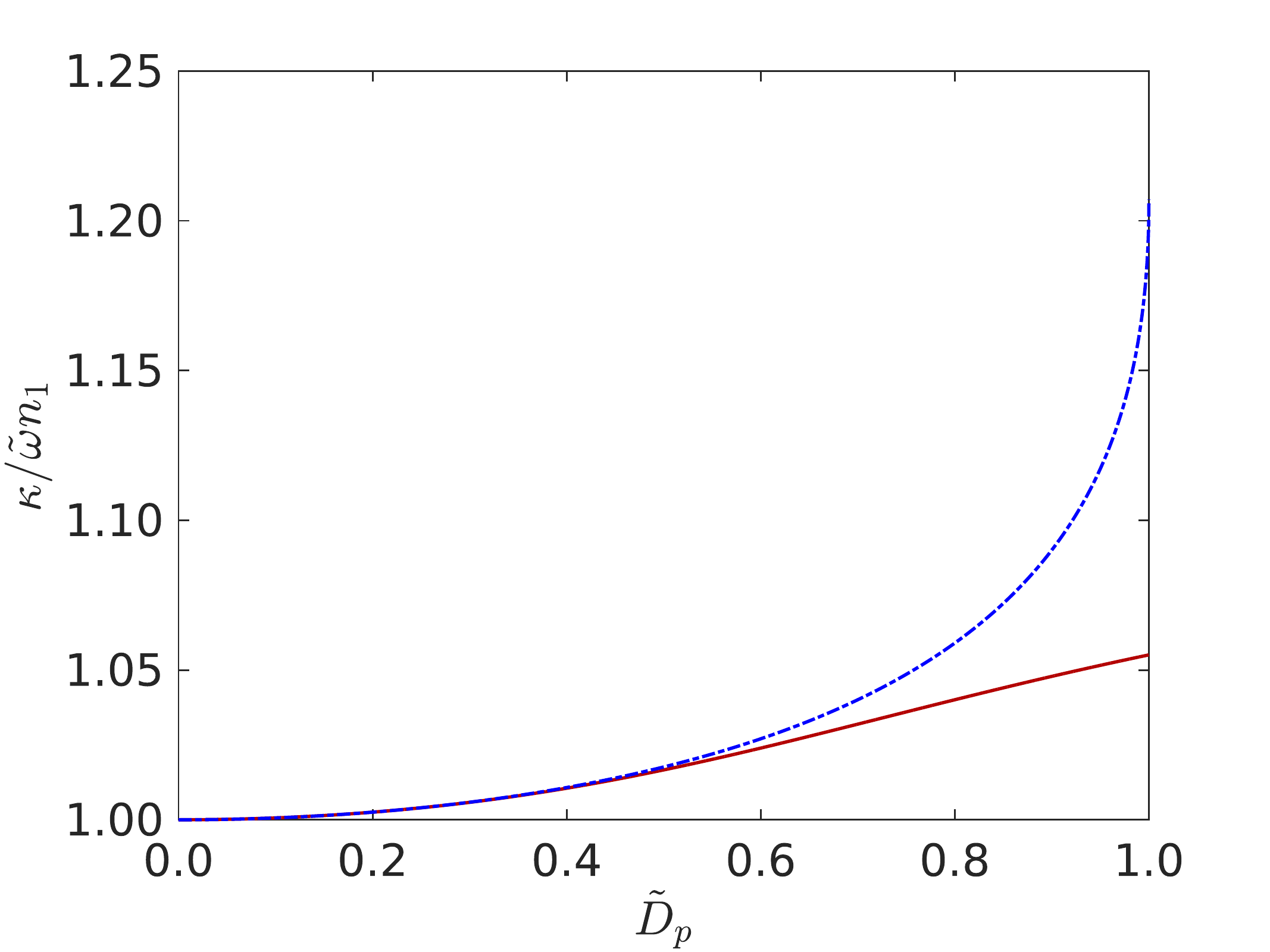}
}

\caption{(Color online) Comparison of exact (solid red curves) and approximate
(blue dash-dot curves) waveguide mode dispersion calculations over
a range of slab thicknesses. The calculation is for a thin uniaxial
slab characterized by $\varepsilon_{layer}^{\parallel}=6.11$ and
$\varepsilon_{layer}^{\perp}=3.03$, suspended in a medium with index
$1.42$, subject to either an $s-$polarized (a) or $p-$polarized
incident field with a vacuum wavelength of $1.55\,{\rm \mu m}$. The
thickness parameters, $\tilde{D}_{s}$ and $\tilde{D}_{p}$ are give
by (\ref{eq:Ds}) and (\ref{eq:Dp}) respectively.}

\label{fig:modes}
\end{figure}

Here we compare the exact and approximate dispersion relations of
the waveguide modes of a thin uniaxial slab. For $\hat{\boldsymbol{z}}$
perpendicular to the slab we take the relative dielectric tensor to
be $\boldsymbol{\varepsilon}_{layer}=\varepsilon_{layer}^{\parallel}\left(\hat{\boldsymbol{x}}\hat{\boldsymbol{x}}+\hat{\boldsymbol{y}}\hat{\boldsymbol{y}}\right)+\varepsilon_{layer}^{\perp}\mathbf{\hat{z}}\mathbf{\hat{z}}$,
with the cladding and substrate of the slab taken to be isotropic
media respectively characterized by relative dielectric constants
$\varepsilon_{1}$ and $\varepsilon_{2}$. The exact solution for
the waveguide modes \cite{Liao2004} of this system is given by
\begin{equation}
\cot\left(hD\right)=\frac{h^{2}-qp}{h\left(q+p\right)},\label{eq:exact_wgcon}
\end{equation}
where 
\begin{align*}
h & =\sqrt{\left(\frac{\varepsilon_{layer}^{\parallel}}{\varepsilon_{layer}^{b}}\right)\left(\tilde{\omega}^{2}\varepsilon_{layer}^{b}-\kappa^{2}\right)},\\
q & =a_{q}\sqrt{\tilde{\omega}^{2}\left(\varepsilon_{layer}^{b}-\varepsilon_{1}\right)-\frac{\varepsilon_{layer}^{b}}{\varepsilon_{layer}^{\parallel}}h^{2}},\\
p & =a_{p}\sqrt{\tilde{\omega}^{2}\left(\varepsilon_{layer}^{b}-\varepsilon_{2}\right)-\frac{\varepsilon_{layer}^{b}}{\varepsilon_{layer}^{\parallel}}h^{2}},
\end{align*}
and where $\varepsilon_{layer}^{b}=\varepsilon_{layer}^{\parallel}$
and $a_{q}=a_{p}=1$ for $s-$polarization, and $\varepsilon_{layer}^{b}=\varepsilon_{layer}^{\perp}$,
$a_{q}=\varepsilon_{layer}^{\parallel}/\varepsilon_{1}$ and $a_{p}=\varepsilon_{layer}^{\parallel}/\varepsilon_{2}$
for $p-$polarization. Taking $\varepsilon_{1}=\varepsilon_{2}$,
approximate dispersion relations can be determined directly from (\ref{eq:mode_conditions});
solving those equations yields

\begin{align}
\frac{\kappa^{2}}{\tilde{\omega}^{2}\varepsilon_{1}} & =1+\frac{1}{4}\tilde{D}_{s}^{2},\label{eq:dispeqs}
\end{align}
for $s-$polarization, and
\begin{equation}
\frac{\kappa^{2}}{\tilde{\omega}^{2}\varepsilon_{1}}=\frac{2}{1+\sqrt{1-\tilde{D}_{p}}},\label{eq:dispeqp}
\end{equation}
for $p-$polarization \footnote{A second formal solution of the second of (\ref{eq:mode_conditions}) is unphysical, in that it corresponds to values of $\kappa^2/\tilde{\omega}\varepsilon_1$ so large that the approximations leading to the derivation of (\ref{eq:mode_conditions}) are invalid.},
where 
\begin{align}
\tilde{D}_{s} & =\tilde{\omega}n_{1}\left(\frac{\varepsilon_{layer}^{\parallel}}{\varepsilon_{1}}-1\right)D,\label{eq:Ds}\\
\tilde{D}_{p} & =\tilde{\omega}n_{1}\left(1-\frac{\varepsilon_{1}}{\varepsilon_{layer}^{\perp}}\right)D.\label{eq:Dp}
\end{align}
The approximate dispersion relations can be shown to agree with the
lowest order solution of the exact relations to first order in the
grating thickness. The exact and approximate dispersion relations
are shown in Fig. \ref{fig:modes} over a range of waveguide thicknesses.
For the purposes of the calculation we set $\varepsilon_{layer}^{\parallel}=6.11$
and $\varepsilon_{layer}^{\perp}=3.03$, with $\varepsilon_{1}=\left(1.42\right)^{2}$,
as used in Fig. \ref{fig:benchplots} with the absence of any diffraction. 

For both polarizations we begin to see significant deviations in Fig.
\ref{fig:modes} as $\tilde{D_{j}}\rightarrow1$, where $j=s,p$.
The $s$-polarized case has a relative deviation of $1.5\%$ at $\tilde{D}_{s}=1$,
which corresponds to a thickness of $86\,{\rm nm}$, while the $p$-polarized
case has a deviation of $27\%$ at $\tilde{D}_{p}=1$. The significantly
larger deviation in the $p$-polarized results can be attributed to
two factors. The first is that for $\tilde{D}_{p}>1$ the square root
in the denominator of (\ref{eq:dispeqs}) becomes imaginary, giving
a firm cut-off for its valid comparison to the exact solution, and
the second is due to the fact that $\varepsilon_{layer}^{\parallel}>\varepsilon_{layer}^{\perp}$
for the cases considered in this paper. While the chosen cut-off for
the s-polarized calculation was found to be $86\,{\rm nm}$, the breakdown
of the p-polarized case occurs at $520\,{\rm nm}$, a thickness well
beyond our underlying assumption that $w_{1}D\ll1$. To provide a
better comparison to the $s-$polarized case, we note that (\ref{eq:dispeqp})
has a relative deviation of approximately $1.5\%$ at $\tilde{D}_{p}=0.7$
which corresponds to a thickness of $364\,{\rm nm}$.

\section{Energy Conservation\label{sec:conserve}}

Here we confirm that our approximate treatment of diffraction and
scattering across the grating satisfies energy conservation exactly
in the limit of no absorption. To do this, we start with the difference
between the total irradiance of the outgoing and incident fields 
\begin{equation}
\begin{array}{rl}
\Delta I= & 2c\epsilon_{0}n_{1}\underset{m}{\sum}\left(\left|\bar{\mathbb{E}}_{out}^{+}\left(\boldsymbol{\kappa}_{m}\right)\right|^{2}+\left|\bar{\mathbb{E}}_{out}^{-}\left(\boldsymbol{\kappa}_{m}\right)\right|^{2}\right)\mathbb{W}\left(\boldsymbol{\kappa}_{m}\right)\cos\theta\left(\boldsymbol{\kappa}_{m}\right)\\
 & -2c\epsilon_{0}n_{1}\underset{m}{\sum}\left(\left|\bar{\mathbb{E}}_{in}^{+}\left(\boldsymbol{\kappa}_{m}\right)\right|^{2}+\left|\bar{\mathbb{E}}_{in}^{-}\left(\boldsymbol{\kappa}_{m}\right)\right|^{2}\right)\mathbb{W}\left(\boldsymbol{\kappa}_{m}\right)\cos\theta\left(\boldsymbol{\kappa}_{m}\right),
\end{array}\label{eq:delIsum}
\end{equation}
where $\mathbb{W}\left(\boldsymbol{\kappa}_{m}\right)\equiv1$ for
propagating fields, and $\mathbb{W}\left(\boldsymbol{\kappa}_{m}\right)\equiv0$
for evanescent fields, such that (\ref{eq:delIsum}) considers the
difference in the incoming and outgoing energy from the grating via
propagating fields; also $\cos\theta\left(\boldsymbol{\kappa}_{m}\right)=w_{1}\left(\boldsymbol{\kappa}_{m}\right)/\tilde{\omega}n_{1}$.
Denoting by $\bar{\theta}$ the diagonal matrix with elements $\bar{\theta}_{mm}=2c\epsilon_{0}n_{1}\cos\theta\left(\boldsymbol{\kappa}_{m}\right)$
we can write the difference in irradiance as 
\begin{equation}
\begin{array}{rl}
\Delta I= & \left[\begin{array}{cc}
\bar{\mathbb{E}}_{out}^{+*} & \bar{\mathbb{E}}_{out}^{-*}\end{array}\right]\left[\begin{array}{cc}
\bar{\mathbb{W}} & \bar{0}\\
\bar{0} & \bar{\mathbb{W}}
\end{array}\right]\left[\begin{array}{cc}
\bar{\theta} & \bar{0}\\
\bar{0} & \bar{\theta}
\end{array}\right]\left[\begin{array}{cc}
\bar{\mathbb{W}} & \bar{0}\\
\bar{0} & \bar{\mathbb{W}}
\end{array}\right]\left[\begin{array}{c}
\bar{\mathbb{E}}_{out}^{+}\\
\bar{\mathbb{E}}_{out}^{-}
\end{array}\right]\\
 & -\left[\begin{array}{cc}
\bar{\mathbb{E}}_{in}^{+*} & \bar{\mathbb{E}}_{in}^{-*}\end{array}\right]\left[\begin{array}{cc}
\bar{\mathbb{W}} & \bar{0}\\
\bar{0} & \bar{\mathbb{W}}
\end{array}\right]\left[\begin{array}{cc}
\bar{\theta} & \bar{0}\\
\bar{0} & \bar{\theta}
\end{array}\right]\left[\begin{array}{cc}
\bar{\mathbb{W}} & \bar{0}\\
\bar{0} & \bar{\mathbb{W}}
\end{array}\right]\left[\begin{array}{c}
\bar{\mathbb{E}}_{in}^{+}\\
\bar{\mathbb{E}}_{in}^{-}
\end{array}\right],
\end{array}\label{eq:delI}
\end{equation}
where $\bar{0}$ is a $2\left(2N+1\right)\times2\left(2N+1\right)$
matrix of zeros, and $\bar{\mathbb{W}}$ is a $2\left(2N+1\right)\times2\left(2N+1\right)$
block diagonal matrix with diagonal elements\textbf{ 
\[
\bar{\mathbb{W}}\left(\boldsymbol{\kappa}_{m}\right)=\mathbb{W}\left(\boldsymbol{\kappa}_{m}\right)\begin{bmatrix}1 & 0\\
0 & 1
\end{bmatrix}.
\]
}After recalling (\ref{eq:S}), (\ref{eq:delI}) becomes 
\begin{align}
\Delta\mathbb{I} & =\left[\begin{array}{cc}
\bar{\mathbb{E}}_{in}^{+*} & \bar{\mathbb{E}}_{in}^{-*}\end{array}\right]\left[\begin{array}{cc}
\bar{\mathbb{W}} & \bar{0}\\
\bar{0} & \bar{\mathbb{W}}
\end{array}\right]\left(\mathscr{S}^{\dagger}\left[\begin{array}{cc}
\bar{\theta}^{\prime} & \bar{0}\\
\bar{0} & \bar{\theta}^{\prime}
\end{array}\right]\mathscr{S}-\left[\begin{array}{cc}
\bar{\theta}^{\prime} & \bar{0}\\
\bar{0} & \bar{\theta}^{\prime}
\end{array}\right]\right)\label{eq:EnCons}\\
 & \cdot\left[\begin{array}{cc}
\bar{\mathbb{W}} & \bar{0}\\
\bar{0} & \bar{\mathbb{W}}
\end{array}\right]\left[\begin{array}{c}
\bar{\mathbb{E}}_{in}^{+}\\
\bar{\mathbb{E}}_{in}^{-}
\end{array}\right],\nonumber 
\end{align}
where $\bar{\theta}^{\prime}=\bar{\mathbb{W}}\bar{\theta}\bar{\mathbb{W}}$
and we have made use of the fact that $\bar{\mathbb{W}}^{2}=\bar{\mathbb{W}}$.

For convenience we introduce the matrix 
\begin{equation}
\bar{\mathcal{C}}^{\pm}=\epsilon_{0}\bar{G}^{\pm}\bar{\chi}_{tot}\left(\bar{1}_{3}-\epsilon_{0}\bar{g}\bar{\chi}_{tot}\right)^{-1},\label{eq:Ctmp}
\end{equation}
where $\bar{\chi}_{tot}=\bar{\chi}_{o}+\bar{\chi}_{v}$. Using (\ref{eq:Ctmp}),
we can write our scattering matrix from (\ref{eq:S}) as 
\begin{equation}
\mathscr{S}=\left[\begin{array}{cc}
\bar{\sigma}_{out}^{+} & \bar{0}\\
\bar{0} & \bar{\sigma}_{out}^{-}
\end{array}\right]\left(\bar{1}_{6}+\left[\begin{array}{cc}
\bar{\mathcal{C}}^{+} & \bar{\mathcal{C}}^{+}\\
\bar{\mathcal{C}}^{-} & \bar{\mathcal{C}}^{-}
\end{array}\right]\right)\left[\begin{array}{cc}
\bar{\sigma}_{in}^{+} & \bar{0}\\
\bar{0} & \bar{\sigma}_{in}^{-}
\end{array}\right],\label{eq:Smat}
\end{equation}
 and can then reduce (\ref{eq:EnCons}) to \begin{widetext} 
\begin{equation}
\begin{array}{rl}
\Delta I= & \left[\begin{array}{cc}
\bar{\mathbb{E}}_{in}^{+*}\bar{\mathbb{W}}\bar{\sigma}_{out}^{+} & \bar{\mathbb{E}}_{in}^{-*}\bar{\mathbb{W}}\bar{\sigma}_{out}^{-}\end{array}\right]\\
 & \cdot\left(\left[\begin{array}{cc}
\bar{\Theta}^{+} & \bar{0}\\
\bar{0} & \bar{\Theta}^{-}
\end{array}\right]\left[\begin{array}{cc}
\bar{\mathcal{C}}^{+} & \bar{\mathcal{C}}^{+}\\
\bar{\mathcal{C}}^{-} & \bar{\mathcal{C}}^{-}
\end{array}\right]+\left[\begin{array}{cc}
\bar{\mathcal{C}}^{+\dagger} & \bar{\mathcal{C}}^{-\dagger}\\
\bar{\mathcal{C}}^{+\dagger} & \bar{\mathcal{C}}^{-\dagger}
\end{array}\right]\left[\begin{array}{cc}
\bar{\Theta}^{+} & \bar{0}\\
\bar{0} & \bar{\Theta}^{-}
\end{array}\right]+\left[\begin{array}{cc}
\bar{\mathcal{C}}^{+\dagger} & \bar{\mathcal{C}}^{-\dagger}\\
\bar{\mathcal{C}}^{+\dagger} & \bar{\mathcal{C}}^{-\dagger}
\end{array}\right]\left[\begin{array}{cc}
\bar{\Theta}^{+} & \bar{0}\\
\bar{0} & \bar{\Theta}^{-}
\end{array}\right]\left[\begin{array}{cc}
\bar{\mathcal{C}}^{+} & \bar{\mathcal{C}}^{+}\\
\bar{\mathcal{C}}^{-} & \bar{\mathcal{C}}^{-}
\end{array}\right]\right)\left[\begin{array}{c}
\bar{\sigma}_{in}^{+}\bar{\mathbb{W}}\bar{\mathbb{E}}_{in}^{+}\\
\bar{\sigma}_{in}^{-}\bar{\mathbb{W}}\bar{\mathbb{E}}_{in}^{-}
\end{array}\right],
\end{array}\label{eq:Entest}
\end{equation}
 \end{widetext} where $\bar{\Theta}^{\pm}=\bar{\sigma}_{in}^{\pm}\bar{\theta}^{\prime}\bar{\sigma}_{out}^{\pm}$.

In simplifying (\ref{eq:Entest}) we find

\begin{align}
\Delta I & =\left[\begin{array}{cc}
\bar{\mathbb{E}}_{in}^{+*} & \bar{\mathbb{E}}_{in}^{-*}\end{array}\right]\begin{bmatrix}\bar{\mathbb{I}}_{++} & \bar{\mathbb{I}}_{+-}\\
\bar{\mathbb{I}}_{-+} & \bar{\mathbb{I}}_{--}
\end{bmatrix}\left[\begin{array}{c}
\bar{\mathbb{E}}_{in}^{+}\\
\bar{\mathbb{E}}_{in}^{-}
\end{array}\right],\label{eq:EnBalance}
\end{align}
where
\begin{align}
\bar{\mathbb{I}}_{\pm\pm}= & \bar{\mathbb{W}}\bar{\sigma}_{out}^{\pm}\left(\bar{\Theta}^{\pm}\bar{\mathcal{C}}^{\pm}+\bar{\mathcal{C}}^{\pm\dagger}\bar{\Theta}^{\pm}+\bar{\mathcal{C}}^{+\dagger}\bar{\Theta}^{+}\bar{\mathcal{C}}^{+}+\bar{\mathcal{C}}^{-\dagger}\bar{\Theta}^{-}\bar{\mathcal{C}}^{-}\right)\bar{\sigma}_{in}^{\pm}\bar{\mathbb{W}},\label{eq:Ieqs}\\
\bar{\mathbb{I}}_{\pm\mp}= & \bar{\mathbb{W}}\bar{\sigma}_{out}^{\pm}\left(\bar{\Theta}^{\pm}\bar{\mathcal{C}}^{\pm}+\bar{\mathcal{C}}^{\mp\dagger}\bar{\Theta}^{\mp}+\bar{\mathcal{C}}^{+\dagger}\bar{\Theta}^{+}\bar{\mathcal{C}}^{+}+\bar{\mathcal{C}}^{-\dagger}\bar{\Theta}^{-}\bar{\mathcal{C}}^{-}\right)\bar{\sigma}_{in}^{\mp}\bar{\mathbb{W}}.\nonumber 
\end{align}

Before further proceeding we note that through the use of the identity
$\bar{\sigma}_{out}^{\pm}\bar{\sigma}_{in}^{\pm}=\bar{1}_{2}$ we
can write 
\begin{equation}
\bar{\Theta}^{\pm}\bar{G}^{\pm}=i\tilde{\omega}cD\bar{\sigma}_{in}^{\pm}\bar{\mathbb{W}}\bar{\sigma}_{out}^{\pm},\label{eq:Gtheta}
\end{equation}
which then allows us to write 
\begin{align}
\bar{\Theta}^{\pm}\bar{\mathcal{C}}^{\pm} & =i\tilde{\omega}\epsilon_{0}cD\bar{\sigma}_{in}^{\pm}\bar{\mathbb{W}}\bar{\sigma}_{out}^{\pm}\bar{\chi}_{tot}\left(\bar{1}_{3}-\epsilon_{0}\bar{g}\bar{\chi}_{tot}\right)^{-1},\label{eq:tc}\\
\bar{\mathcal{C}}^{\pm\dagger}\bar{\Theta}^{\pm} & =-i\tilde{\omega}\epsilon_{0}cD\left(\bar{1}_{3}-\epsilon_{0}\bar{\chi}_{tot}^{\dagger}\bar{g}^{\dagger}\right)^{-1}\bar{\chi}_{tot}^{\dagger}\bar{\sigma}_{in}^{\pm}\bar{\mathbb{W}}\bar{\sigma}_{out}^{\pm},\nonumber 
\end{align}
where in the latter expression we have made use of the fact that $\left(\bar{\sigma}_{in}^{\pm}\bar{\mathbb{W}}\bar{\sigma}_{out}^{\pm}\right)^{\dagger}=\bar{\sigma}_{in}^{\pm}\bar{\mathbb{W}}\bar{\sigma}_{out}^{\pm}$.
Next, note that when we premultiply the $3\left(2N+1\right)\times3\left(2N+1\right)$
block diagonal matrix $\bar{G}^{+}$ by the $3\left(2N+1\right)\times3\left(2N+1\right)$
block diagonal matrix $\bar{\sigma}_{in}^{+}\bar{\mathbb{W}}\bar{\sigma}_{out}^{+}$
we get a block diagonal matrix $\bar{\sigma}_{in}^{+}\bar{\mathbb{W}}\bar{\sigma}_{out}^{+}\bar{G}^{+}$
in which each block associated with a propagating (diffracted or scattered)
order equals the corresponding block of $\bar{G}^{+}$, but which
vanishes if the block is associated with an evanescent order. In the
same way the blocks of $\bar{\sigma}_{in}^{-}\bar{\mathbb{W}}\bar{\sigma}_{out}^{-}\bar{G}^{-}$
associated with propagating orders equal the corresponding blocks
of $\bar{G}^{-}$, while those associated with evanescent orders vanish.
Recalling that $\bar{g}\left(\boldsymbol{\kappa}_{m}\right)=\frac{1}{2}\left[\bar{G}^{+}\left(\boldsymbol{\kappa}_{m}\right)+\bar{G}^{-}\left(\boldsymbol{\kappa}_{m}\right)\right]$,
we have $\bar{\sigma}_{in}^{+}\bar{\mathbb{W}}\bar{\sigma}_{out}^{+}\bar{G}^{+}+\bar{\sigma}_{in}^{-}\bar{\mathbb{W}}\bar{\sigma}_{out}^{-}\bar{G}^{-}=\bar{g}-\bar{g}^{\dagger}$,
where the sum on the left hand side will either equal $2\bar{g}$
or $\bar{0}$ as $\bar{g}\left(\boldsymbol{\kappa}_{m}\right)=-\bar{g}^{\dagger}\left(\boldsymbol{\kappa}_{m}\right)$
for propagating orders, while $\bar{g}\left(\boldsymbol{\kappa}_{m}\right)=\bar{g}^{\dagger}\left(\boldsymbol{\kappa}_{m}\right)$
for evanescent orders. With this in mind and from the use of (\ref{eq:tc})
we have

\begin{equation}
\begin{array}{rl}
 & \bar{\mathcal{C}}^{+\dagger}\bar{\Theta}^{+}\bar{\mathcal{C}}^{+}+\bar{\mathcal{C}}^{-\dagger}\bar{\Theta}^{-}\bar{\mathcal{C}}^{-}=\\
 & -i\tilde{\omega}\epsilon_{0}^{2}cD\left(\bar{1}_{3}-\epsilon_{0}\bar{\chi}_{tot}^{\dagger}\bar{g}^{\dagger}\right)^{-1}\bar{\chi}_{tot}^{\dagger}\left(\bar{g}-\bar{g}^{\dagger}\right)\bar{\chi}_{tot}\left(\bar{1}_{3}-\epsilon_{0}\bar{g}\bar{\chi}_{tot}\right)^{-1}.
\end{array}\label{eq:RthetaR}
\end{equation}
 Finally, through the substitution of the expressions for $\bar{\Theta}^{\pm}\bar{\mathcal{C}}^{\pm}$,
$\bar{\mathcal{C}}^{\pm\dagger}\bar{\Theta}^{\pm}$, (\ref{eq:tc},\ref{eq:RthetaR})
into (\ref{eq:Ieqs}), and some simple algebra, we find $\bar{\mathbb{I}}_{\pm\pm}=\bar{\Gamma}^{\pm}\bar{\sigma}_{in}^{\pm}\bar{\mathbb{W}}$
and $\bar{\mathbb{I}}_{\pm\mp}=\bar{\Gamma}^{\pm}\bar{\sigma}_{in}^{\mp}\bar{\mathbb{W}}$
where
\begin{align}
\bar{\Gamma}^{\pm} & =i\tilde{\omega}\epsilon_{0}cD\bar{\mathbb{W}}\bar{\sigma}_{out}^{\pm}\left(\bar{1}_{3}-\epsilon_{0}\bar{\chi}_{tot}^{\dagger}\bar{g}^{\dagger}\right)^{-1}\label{eq:Idiff}\\
 & \cdot\left[\bar{\chi}_{tot}-\bar{\chi}_{tot}^{\dagger}\right]\left(\bar{1}_{3}-\epsilon_{0}\bar{g}\bar{\chi}_{tot}\right)^{-1}.\nonumber 
\end{align}
For non-absorbing gratings $\bar{\chi}_{tot}$ is Hermitian, $\bar{\mathbb{I}}_{\pm\pm}=\bar{\mathbb{I}}_{\pm\mp}=0$,
and $\Delta I=0$ from (\ref{eq:EnBalance}). If absorption is present
(\ref{eq:Idiff}) and (\ref{eq:EnBalance}) can be used to calculate
its effect on the energy balance between incident and scattered fields. 

\bibliographystyle{apsrev4-1}
\bibliography{green-grating}

\end{document}